\documentclass{iopjournal}

\usepackage[square,numbers,sort&compress]{natbib}
\usepackage{graphicx}
\expandafter\let\csname equation*\endcsname\relax
\expandafter\let\csname endequation*\endcsname\relax
\usepackage{amsmath,amsfonts,amssymb}
\usepackage{subcaption}
\usepackage{placeins}

\hypersetup{hypertexnames=false}


\begin{document}

\articletype{Paper} 

\title{Numerical error analysis of EMC3-EIRENE
simulations in the HSX stellarator}

\author{F.A. Chavarria$^{1,*}$\orcid{0009-0008-9106-1058}, D. Boeyaert$^1$\orcid{0000-0003-0920-8660}, H. Frerichs$^1$\orcid{0000-0002-3527-5106}, K. A. Garcia$^2$\orcid{0009-0007-3332-1563}, B. Geiger$^1$\orcid{0000-0001-8706-1874}}

\affil{$^1$Department of Nuclear Engineering and Engineering Physics, University of Wisconsin--Madison, Madison, WI 53706, United States of America}

\affil{$^2$University of Greifswald, Greifswald, Germany}

\affil{$^*$Author to whom any correspondence should be addressed.}

\email{fchavarria@wisc.edu}


\begin{abstract}
Plasma edge simulations with codes like EMC3-EIRENE are widely used to study fusion magnetic confinement devices. However, numerical errors can significantly impact the simulation solution and are often insufficiently characterized. These errors consist of the statistical error, bias, and time integration error, resulting from the finite sampling and tracking of Monte Carlo particles, and the discretization error, resulting from a finite grid resolution. This work presents the first comprehensive numerical error analysis of EMC3-EIRENE simulations in stellarator geometries, focusing on edge plasmas in the Helically Symmetric eXperiment (HSX). Fundamental plasma and neutral quantities are analyzed, and error reduction rates are compared against theoretical expectations. The statistical error, bias, and time integration error are fully characterized; however, quantification of the discretization error remains challenging because of the complex grid geometries used in EMC3-EIRENE. The commonly used volume averaged relative change metric is shown to systematically underestimate the statistical error, highlighting a critical flaw in EMC3-EIRENE analysis, where convergence is claimed despite significant noise remaining in the simulation solution. Dominant contributions of the time integration error are shown to differ in stellarator simulations compared to tokamak simulations, related to the relative importance of cross-field transport between the two devices. Based on these results and existing literature, we establish best practices towards improved accuracy and efficiency in general stellarator modeling with EMC3-EIRENE.
\end{abstract}


\section{Introduction}
\label{sec:introduction}

In fusion research, understanding plasma behavior near plasma facing components (PFCs) is extremely important, both for furthering understanding and performance of present day experiments, and for guiding the design of future devices. However, the plasma edge is complex due to the interaction of plasma, atomic, and molecular processes, where diagnostic capabilities are often limited. Therefore, numerical simulations are needed to analyze the physics of this region. In non-axisymmetric devices like stellarators, the plasma and neutral transport near the edge of 3D devices is often simulated with the EMC3-EIRENE code \cite{feng3DEdgeModeling2004, reiterEIRENEB2EIRENECodes2005}.

The 3D coupled Monte Carlo--Monte Carlo (MC--MC) code EMC3-EIRENE is commonly used to study the scrape-off layer (SOL) in non-axisymmetric fusion devices. It has been used to study stellarators such as W7-AS, W7-X, LHD, CTH, and HSX \cite{fengTransportIslandDivertors2002, wintersEMC3EIRENESimulationFirst2021, boeyaertAnalysisNeutralFluxes2023, fengComparativeDivertortransportStudy2009, garciaExplorationNonresonantDivertor2023, baderSimulationsEdgeConfigurations2013}, as well as tokamaks with symmetry breaking resonant magnetic perturbations such as DIII-D and ITER \cite{frerichsImpactScreeningResonant2012, schmitzThreedimensionalModelingPlasma2016, vanblarcumComparisonDivertorHeat2026}. EMC3 solves a simplified, time-independent, version of the Braginskii fluid equations with a MC technique \cite{feng3DFluidModelling1999}. This technique is adopted in EMC3 to enable a clean separation of parallel and perpendicular transport in strongly anisotropic plasmas, thereby avoiding artificial cross-field diffusion errors that can arise in finite volume methods when applied to complex 3D magnetic topologies \cite{feng3DEdgeModeling2004, fengSimpleHighlyAccurate2005}. EIRENE similarly uses a MC technique to solve a Boltzmann-type equation for neutral particle interactions in position--velocity phase space \cite{reiterEIRENEB2EIRENECodes2005}. These two codes are strongly coupled and solved iteratively by alternately updating the plasma and neutral solutions, where each step utilizes the results of the previous iterations, until a steady-state solution is reached.  

It is important to properly quantify the error in codes such as EMC3-EIRENE to ensure an accurate solution. Four types of errors are present in a simulation: code, user, modeling, and numerical errors \cite{oberkampfErrorUncertaintyModeling2002}. Code and user errors, which consist of human mistakes when writing and using a model, respectively, although important, are not the focus of the presented work. Modeling errors are defined as the difference between the physical solution and the exact solution of the model equations \cite{ghoosAccuracybasedSimulationStrategies2019}, and are the result of nonphysical assumptions taken to solve the complex system. One such example is the omission of drifts in EMC3 \cite{dewolfPlasmaDriftsEMC32024, wolfImplementationExBGradB2026}. Modeling errors are not studied further in this paper. Numerical errors consist of inaccuracies introduced during the calculation process due to the finite nature of computational arithmetic, and are the difference between the exact solution of the model equations and the solution obtained with the code. They are closely linked with the implemented numerical strategy. In EMC3-EIRENE simulations, the numerical error $\epsilon_{num}$ is composed of four parts \cite{ghoosNumericalAccuracyConvergence2020}:
\begin{equation}
    \epsilon_{num} = \epsilon_s + \epsilon_b + \epsilon_t + \epsilon_d
    \label{eq:num_error}
\end{equation}
The statistical error $\epsilon_s$ appears as statistical noise associated with the finite sampling of MC particles and is described by a probability distribution with mean zero. The bias $\epsilon_b$ is the deterministic error arising from the same finite sampling of MC particles. The time integration error $\epsilon_t$ results from advancing fluid MC particles through discrete time steps corresponding to random walks. The discretization error $\epsilon_d$ originates from solving continuous equations at finite points in space and is often the dominant error in simulations because its reduction requires significantly greater computational resources compared to the other errors \cite{boeyaertNumericalErrorAnalysis2022, ghoosGridResolutionStudy2018}. Quantifying and reducing numerical errors allows the simulation solution to be properly interpreted. In addition to these errors, modeling uncertainties, such as uncertainties and limitations in the atomic and molecular rate coefficients used in simulations, remain and will also influence the simulation accuracy \cite{cappelliAnalysisModelingTungsten2026}. Modeling uncertainties are not studied further in this paper.

The inherent stochastic nature of EMC3-EIRENE creates a multi-layered numerical error structure, which significantly complicates the analysis of accuracy and convergence. For the 2D Finite Volume--Monte Carlo code SOLPS-ITER \cite{wiesenNewSOLPSITERCode2015}, comprehensive numerical error analysis techniques have been developed in \cite{ghoosAccuracybasedSimulationStrategies2019,baelmansEfficientCodeSimulation2017, baetenAnalyticalStudyStatistical2018}, and have been successfully implemented to quantify error in simulation solutions \cite{ghoosGridResolutionStudy2018, boeyaertNumericalErrorAnalysis2022, uytvenDiscretizationErrorEstimation}. This was further applied in \cite{ghoosNumericalAccuracyConvergence2020} to optimize the balance between accuracy and computational time in EMC3-EIRENE for tokamak simulations. However, the systematic quantification of numerical accuracy in EMC3-EIRENE remains an outstanding challenge, particularly within complex stellarator geometries. This topic is especially relevant as EMC3-EIRENE often suffers from memory restraints for large scale research and reactor-relevant devices, thus necessitating a balance of computational requirements and numerical accuracy.



This work applies the methodology established in \cite{ghoosAccuracybasedSimulationStrategies2019} to the HSX stellarator, but is generally applicable to all EMC3-EIRENE modeling. In comparison with \cite{ghoosNumericalAccuracyConvergence2020}, this work provides error estimations for key plasma and neutral quantities, with particular emphasis on the error's effect through the simulation domain. In addition, the discretization error is formally described as a whole for the first time, and initial measurements are established.

The paper is organized as follows. In the next section, we describe the setup and model of the simulations. In section \ref{sec:numerical_errors}, we describe the numerical errors present in EMC3-EIRENE, with emphasis on the complications of the discretization error. In section \ref{sec:Simulation_approach}, the implemented iteration procedure and the data acquisition are discussed. In section \ref{sec:results}, we provide estimates of the statistical error, bias, time integration error, and initial estimates of the discretization error. The main conclusions are summarized in section \ref{sec:conclusion}, and the paper concludes with some general recommendations for numerical setups in EMC3-EIRENE.


\section{Simulation setup}
\label{sec:simulation_setup}

Simulations are performed for the Helically Symmetric eXperiment (HSX) \cite{andersonHelicallySymmetricExperiment1995, Geiger_HSX_2026}. The HSX stellarator, located at the University of Wisconsin-Madison, features a major radius of 1.2~m, a plasma volume of 0.44~m$^3$, and a magnetic field up to 1~T. Due to its unique quasi-helically symmetric (QHS) magnetic field and an edge topology consistent with the characteristics of non-resonant divertors \cite{baderHSXExampleResilient2017, garciaResilientStellaratorDivertor2025, boeyaertImprovedNeutralExhaust2025}, HSX serves as a vital platform for validating the fundamental physics required for future stellarator designs.

\begin{figure}
\centering
\includegraphics[width=0.49\textwidth]{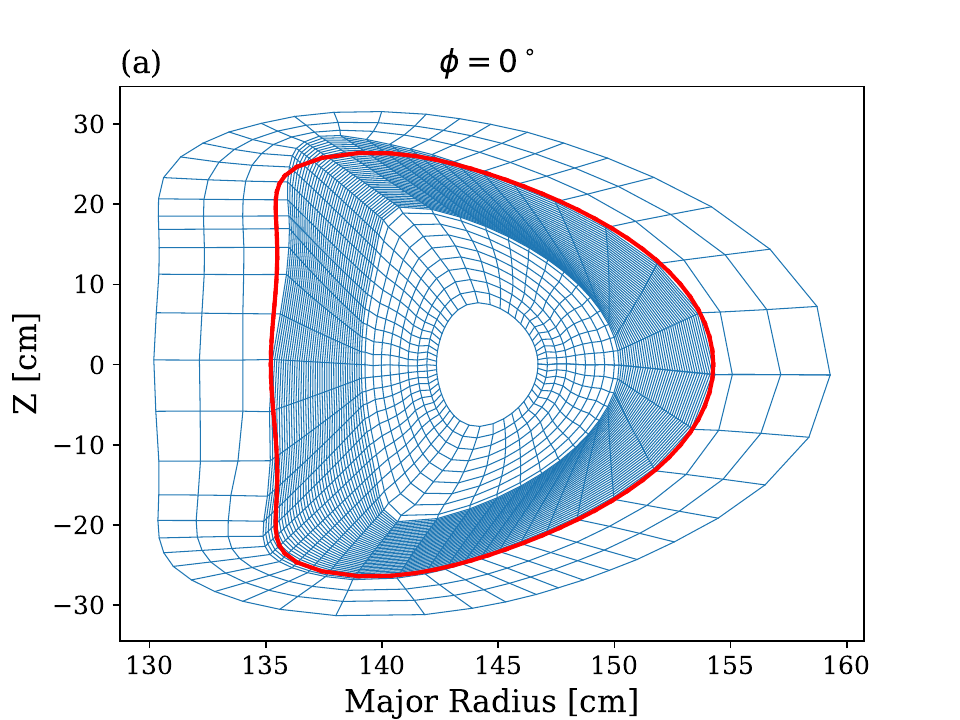}
\hfill
\includegraphics[width=0.49\textwidth]{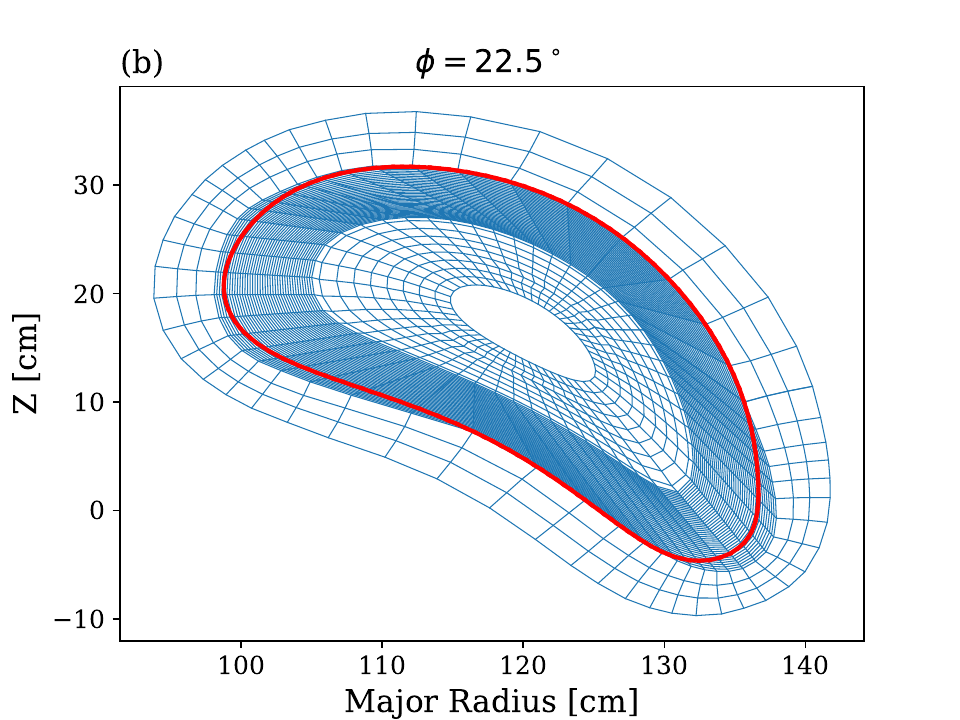}
\caption{Magnetic mesh for HSX's QHS geometry at different toroidal angles (vessel wall in red): (a) $\phi=0^\circ$, and (b) $\phi=22.5^\circ$. Note that this is not the mesh used in this study but rather a coarse representation of EMC3-EIRENE meshes for HSX for better visualization.}
\label{fig:mesh}
\end{figure}

HSX exhibits four-fold symmetry in addition to stellarator symmetry, meaning every half-field period is an up-down mirror of the other half. Therefore, it is sufficient to model only $45^\circ$ of the device. EMC3-EIRENE requires a grid consisting of two domains: a field-aligned plasma domain, and an extended neutral domain used to calculate atomic/molecular reactions and plasma surface interactions further into the core and regions far away from the plasma (such as the sub-divertor structures for devices such as W7-X \cite{boeyaertAnalysisNeutralFluxes2023}). The field-aligned domain encompasses both open field lines that terminate at PFCs and closed field lines within the core and islands. The field-aligned grid relies on three coordinate directions: a radial direction $r$ bounded by inner and outer contours, a poloidal direction $\theta$, and a field-aligned direction $\phi$. The magnetic field needed for grid construction is derived from the QHS vacuum field by solving the Biot-Savart equations for the main non-planar coils of HSX using the MAKEGRID code from the STELLOPT package \cite{doecode_12551}. Using the FLARE code \cite{frerichsFLAREFieldLine2024}, the 3D grid is constructed by first initializing a 2D base mesh at the $\phi=22.5^\circ$ plane. For HSX, the 2D base mesh is radially extended until it intersects the vessel wall as shown in figure~\ref{fig:mesh}. This defines the spatial extent of the field-aligned domain. Field lines are then tracked bi-directionally to define subsequent toroidal planes through the device. The extended neutral mesh is created by radially extending the field-aligned mesh further into the core and vacuum regions. EMC3 will only solve for quantities within the field-aligned section of the grid, and EIRENE will operate on the entire grid. The primary grid of this study has a resolution of $150 \times 400 \times 180$ in the radial, poloidal, and field-aligned directions, respectively. Two 2D R-Z slices of an example HSX mesh are provided in figure \ref{fig:mesh}.

A requirement of EMC3-EIRENE is that the $\phi=0^\circ$ and $\phi=45^\circ$ toroidal planes satisfy up-down symmetry, which means that the corresponding outermost boundaries of the extended neutral domain are symmetric with respect to the midplane defined as $Z=0$. A reversible field line mapping (RFLM) technique, which preserves stellarator symmetry, performs the coordinate transformation across this boundary \cite{fengSimpleHighlyAccurate2005}. Typically, it is customary to divide the simulation domain into further toroidal zones to maintain adequate flux conservation in regions of high magnetic shear or large radial excursions of field lines \cite{frerichsBlockstructuredGridsLagrangian2010, baderSimulationsEdgeConfigurations2013}, but as this can create additional errors in the solution, it is not done here. However, this does create cell distortions in the extended neutral mesh. Therefore, we extend the neutral mesh significantly further than the field-aligned grid to force up-down symmetry at the outermost boundary to be satisfied. This is appropriate for HSX as nearly the entirety of the SOL in HSX is encompassed by the field-aligned grid as seen in figure~\ref{fig:mesh}. Grid cells (tracking volumetric quantities) are made progressively larger away from the vessel wall through field line tracing methods, preserving fine resolution near the vessel wall. These last two points are closely related to the error in the simulation solution and will be further discussed in subsection \ref{subsec:discretization_error}.

EMC3-EIRENE is divided into three modules that are solved iteratively: the ENERGY module, the STREAMING module, and the NEUTRAL (EIRENE) module \cite{frerichsVolumetricRecombinationEMC3EIRENE2021}. The ENERGY module solves the electron and ion energy equations for the electron temperature $T_e$ and ion temperature $T_i$, the STREAMING module solves the continuity and momentum equations for the plasma density $n_i$ and Mach number $M$, and the NEUTRAL module solves for neutral source terms arising from plasma-neutral interactions. Similar to SOLPS-ITER, EMC3-EIRENE is under-relaxed, which limits change in the solution iteration-to-iteration and ensures that unsteady terms vanish asymptotically under steady-state conditions \cite{ghoosAccuracybasedSimulationStrategies2019}. In these simulations, the relaxation factor is taken as 0.5, 0.4, and 0.4 for the ENERGY, STREAMING, and NEUTRAL modules, respectively. These are default values often taken in ECM3-EIRENE \cite{frerichsStabilityControlIterated2015}.

We select a representative plasma condition applicable to larger devices by basing the imposed density and input power on future operating parameters after an ongoing electron cyclotron resonance heating (ECRH) system upgrade \cite{APS_HSXUpgrades}. This future system uses a new 70 GHz gyrotron from W7-AS which allows for a total injected power up to 300 kW, and a target maximum density of $3 \times 10^{13}$ cm$^{-3}$ on-axis. The density is fixed at the innermost boundary of EMC3's field-aligned mesh (roughly $r/a =0.75$) at $2 \times 10^{13}$ cm$^{-3}$. A total of 200 kW enters the plasma edge from the core and we take the assumption that it is divided equally among ions and electrons. Although impurities are inherently present in experimental devices and have been shown to significantly influence plasma transport in the edge \cite{briesemeisterFlowVelocityMeasurements2010}, only hydrogen is simulated in this work. The cross-field transport coefficients are taken uniformly across the computational domain, with $D=1$ m$^2$/s for the ion diffusion and $\chi_e=\chi_i=3$ m$^2$/s for the electron and ion thermal diffusivity. These are customary values often used in stellarator simulations \cite{boeyaertAnalysisNeutralFluxes2023, wintersEMC3EIRENESimulationFirst2021}. At the PFCs, the Bohm criterium is applied, requiring plasma speed to reach the sound speed.



\section{Numerical errors in EMC3-EIRENE}
\label{sec:numerical_errors}

The coupling of the ENERGY, STREAMING, and NEUTRAL modules in EMC3-EIRENE means that the total numerical error $\epsilon_{total\,num}$ considers the sum of the numerical errors from each of these three modules \cite{ghoosNumericalAccuracyConvergence2020}:
\begin{equation}
    \epsilon_{total\,num} = \epsilon_{ENE} + \epsilon_{STR} + \epsilon_{NEU}
    \label{eq:module_error}
\end{equation}
The errors of the ENERGY and STREAMING modules ($\epsilon_{ENE}$ and $\epsilon_{STR}$) consist of all contributions in equation (\ref{eq:num_error}), while the errors of the NEUTRAL module ($\epsilon_{NEU}$) consist of only the statistical error, bias, and discretization error. This is attributed to EIRENE using a kinetic model, compared to the fluid model of EMC3. Although each module produces its own discretization error, individual contributions cannot be easily isolated due to the iteration procedure implemented in the code. Therefore, the discretization error is analyzed as the combined effect of all modules, representative of the EMC3-EIRENE model as a whole. The remainder of this section describes the errors in equation (\ref{eq:num_error}) and the methods used to characterize them.

\subsection{Statistical error}
\label{subsec:stat_error}

The statistical error in EMC3-EIRENE has a probability distribution with mean zero and standard deviation $\sigma$. It is estimated using the central limit theorem as
\begin{equation}
    \epsilon_s \approx \sigma = \frac{\sigma_1}{\sqrt{\frac{I}{\tau}}} = \frac{A_s(h) \sqrt{\tau}}{\sqrt{PI}}
    \label{eq:stat_error}
\end{equation}
where $\sigma_1$ ($\sigma_1 = A_s(h)/\sqrt{P}$) is the standard deviation of iterations (calculated as the sample standard deviation), $A_s$ characterizes the magnitude of the error, $h$ is some measure of grid spacing, $P$ is the number of simulated MC particles, and $I$ is the number of iterations the solution is averaged over \cite{ghoosNumericalAccuracyConvergence2020}. $A_s$ is explicitly expressed as a function of $h$ as the statistical error depends on the spatial discretization. In practice, this means that a better spatial resolution is penalized by a larger variance in the solution \cite{xuAssessmentNumericalAccuracy1999}. Furthermore, it has been shown in \cite{ghoosNumericalAccuracyConvergence2020} and verified in this study (subsection \ref{subsec:results_time_integration_error}) that $\epsilon_s$ has no significant dependence on the temporal discretization.

The correlation time $\tau$ in equation (\ref{eq:stat_error}) measures the dependency iteration-to-iteration in the simulation and should be interpreted as the number of iterations required until a statistically independent solution is produced. It is defined as
\begin{equation}
    \tau = 1 + 2\sum_{\eta=1}^{\infty}\rho(\eta)
    \label{eq:auto_correlation}
\end{equation}
with $\rho(\eta)$ the normalized autocorrelation with lag $\eta$ \cite{ghoosNumericalErrorEstimation2017, NonlinearTimeSeries2003}. In this study, two methods are used to measure the correlation time: the batching method and the autocorrelation method. For the batching method, we divide consecutive iterations into batches of size $N$. The calculation of $\tau$ is then
\begin{equation}
    \tau = \frac{\sigma_N^2}{\sigma_1^2}N
    \label{eq:batching}
\end{equation}
where $\sigma_1^2$ is the sample variance of iterations and $\sigma_N^2$ is the sample variance of the batch means. The batches can be non-overlapping, where each batch starts where the previous batch ended, or overlapping, where the next batch begins at each new iteration. Both the non-overlapping and overlapping batching methods are used in this study. With the autocorrelation method, we fit $\rho(\eta)$ in the form of an exponential decaying function and compute $\tau$ with the corresponding fit \cite{newmanMonteCarloMethods1999}. Further information on applying these two methods to plasma edge simulations can be found in \cite{ghoosNumericalErrorEstimation2017}.

As EMC3-EIRENE has three modules, a statistical error will result from the MC particle population of each module. However, the statistical error only directly affects variables calculated within the module of origin. Errors propagating from external modules influence the variable in an average sense, manifesting through the bias \cite{ghoosNumericalAccuracyConvergence2020}. As guided in \cite{ghoosGridResolutionStudy2018}, the estimate of the statistical error is multiplied by three to account for inaccuracies in the calculation of $\tau$ and $\sigma_1$ due to the stochastic nature of the statistical error, and the limitation on iterations available to us.

\subsubsection{Standard measures of statistical error}
\label{subsubsec:standard_stat_error}

Although equation (\ref{eq:stat_error}) is a good measure of the statistical error, it is not the standard measure used by EMC3-EIRENE users. Instead, modelers use the volume averaged relative change metric \cite{kobayshEMC3-EIRENEaccuracyconvergence}. It is defined as
\begin{equation}
    \Delta f^{(n)} = \frac{\sum_j\left|f_j^{(n-1)}-f_j^{(n)}\right|V_j}{\sum_j\left(f_j^{(n-1)}+f_j^{(n)}\right)V_j}
    \label{eq:relative_change}
\end{equation}
where $n$ is the iteration index, $j$ is the index for a grid cell of volume $V_j$, and $f$ is the solution field (typically $n_i$, $T_e$, or $T_i$). $\Delta f^{(n)}$ roughly scales $\propto~1/\sqrt{P}$ \cite{frerichsBlockstructuredGridsLagrangian2010}, but it does not account for iteration-to-iteration dependency, averaging, or distance from the mean solution. Therefore, it is stressed that although equation (\ref{eq:relative_change}) is computational inexpensive, it is not a rigorous metric of the statistical error and a metric based on the standard deviation (i.e. equation (\ref{eq:stat_error})) is preferred. This point will be discussed further in subsection \ref{subsec:results_statistical_error}.

\subsection{Bias}
\label{subsec:bias}

The bias is the deterministic error caused by using a finite number of MC particles \cite{xuAssessmentNumericalAccuracy1999}. It is composed of the convergence error due to residuals remaining in the solution and the finite sampling bias resulting from nonlinearities in the code. The bias is inversely proportional to the simulated MC particle count $\epsilon_b \propto 1/P$, and, similar to the statistical error, a finer grid resolution is penalized by a larger bias. We estimate the bias using Richardson-like extrapolation \cite{ghoosAccuracyConvergenceCoupled2016}, where we compare two simulations with MC particle counts that differ by a constant $\alpha$ ($P$ and $\alpha P$) on the same grid and estimate the error as
\begin{equation}
    \epsilon_{b} \approx \frac{f_{P} - f_{\alpha P}}{1-\alpha^{p_b}}
    \label{eq:bias_richardson}
\end{equation}
where $f_P$ is a volumetric quantity obtained with $P$ simulated MC particles, and $p_b$ is the order of convergence of the bias. From the asymptotic scaling of $\epsilon_b$, we expect $p_b=-1$. Typically, $p_b$ is directly calculated with a third simulation following $\alpha^2 P$ MC particles \cite{boeyaertNumericalErrorAnalysis2022}. However, this estimate is sensitive to statistical noise, rendering it impractical for EMC3-EIRENE. Instead, $p_b$ is calculated by observing the scaling of the bias across many simulations (subsection \ref{subsec:results_bias}). As both $\epsilon_s$ and $\epsilon_b$ depend on $P$, to ensure a good estimate of $\epsilon_b$, averaging should be performed to guarantee that $\epsilon_s < \epsilon_b$ \cite{ghoosNumericalErrorEstimation2017}. 

As discussed in subsection \ref{subsec:stat_error}, a bias will result from the MC particle population of each EMC3-EIRENE module. Furthermore, since equation (\ref{eq:bias_richardson}) assumes that the difference between $f_P$ and $f_{\alpha P}$ is exclusively from the change in $P$, the bias resulting from each module will be analyzed individually. Due to the remaining statistical error in the solution, and since the theoretical scaling of the bias may not yet be reached, a safety factor of two is applied to the estimated bias \cite{ghoosGridResolutionStudy2018}. 

\subsection{Time integration error}
\label{subsec:time_integration_error}

EMC3's transport equations obey a generalized conservation principle for density, velocity, and ion and electron temperatures. This allows the creation of one general numerical method usable for determining $n_i$, $M$, $T_e$, and $T_i$ despite their different boundary conditions \cite{feng3DFluidModelling1999}. The transport equations are solved by first generating MC particles according to a given source distribution, and then following them through small time steps $\Delta t$, with the corresponding random walk step $\Delta r$ determined by
\begin{equation}
    \Delta r = \sqrt{2\beta_{\parallel}\Delta t_\parallel}\xi_\parallel  + \alpha_{\parallel}\Delta t_\parallel + \sqrt{2\beta_{\perp}\Delta t_\perp}\xi_\perp  + \alpha_{\perp}\Delta t_\perp
    \label{eq:random_walk}
\end{equation}
where $\beta_{\parallel}$, $\alpha_{\parallel}$, $\beta_{\perp}$, and $\alpha_{\perp}$ represent transport coefficients for diffusive ($\beta$) and convective ($\alpha$) processes, and $\xi_\parallel$ and $\xi_\perp$ are random unit vectors parallel and perpendicular to the magnetic field, respectively. The transport coefficients depend on which conservation principle is being simulated (mass, momentum, or energy), and are determined by rewriting the model equations in Fokker-Planck form \cite{feng3DFluidModelling1999}. The dependence of $\Delta r$ on $\Delta t$ creates the time integration error.

This stochastic random walk process is known as Euler--Maruyama discretization \cite{kloedenNumericalSolutionStochastic1992}. As a result, macroscopic plasma parameters converge linearly with respect to the chosen time step $\epsilon_t \propto \Delta t$ (i.e. $p_t=1$ where $p_t$ is the convergence order of the time integration error) \cite{ghoosNumericalAccuracyConvergence2020}. The high anisotropy of energy transport means that in practice the perpendicular ENERGY time step is often chosen to be roughly an order of magnitude larger than the parallel ENERGY time step ($\Delta t_{E\perp} > \Delta t_{E\parallel}$) to speed up the simulation time at low cost to accuracy. Conversely, as particle transport is generally less anisotropic than energy transport, the STREAMING module has only one time step value $\Delta t_{S}$ corresponding to transport in both directions.

In this paper, we follow the approach of \cite{ghoosNumericalAccuracyConvergence2020} where we characterize the error as a function of $\Delta t$, rather than resulting from the non-uniform transport coefficients. The time integration error is deterministic, indicating we can also utilize Richardson-like extrapolation to estimate it. Therefore, we use two simulations with different time steps ($\Delta t$ and $\alpha \Delta t$) on the same grid, and estimate the error $\epsilon_t$ with an calculation equivalent to equation (\ref{eq:bias_richardson}). Similar to the bias, $p_t$ will be calculated by measuring the time integration error across multiple simulations (subsection~\ref{subsec:results_time_integration_error}), all three time steps will be analyzed individually, and a safety factor of two is applied to the estimated time integration error, accounting for remaining statistical errors and the fact that the theoretical scaling may not yet be reached. 

\subsection{Discretization error}
\label{subsec:discretization_error}

The discretization error arises from approximations related to solving continuous equations on a finite number of grid cells. As discussed in section \ref{sec:simulation_setup}, the EMC3-EIRENE grid consists of the primary field-aligned grid and the extended neutral grid. Both sections will create a discretization error, but the emphasis on this analysis will be on the field-aligned grid. Richardson extrapolation \cite{roachePerspectiveMethodUniform1994, roacheQuantificationUncertaintyComputational1997} is a common method used to analyze the discretization error in plasma edge codes \cite{ghoosGridResolutionStudy2018, boeyaertNumericalErrorAnalysis2022, uytvenDiscretizationErrorEstimation}. However, despite its previous use in 3D tokamak simulations \cite{ghoosNumericalAccuracyConvergence2020}, Richardson extrapolation cannot fully describe the discretization error in EMC3-EIRENE. Consequently, this study focuses exclusively on the radial and poloidal discretization error. The motivation for this shift and the theoretical framework for analyzing the discretization error are detailed in the following paragraphs.

\begin{figure}
\centering
\includegraphics[width=0.49\textwidth]{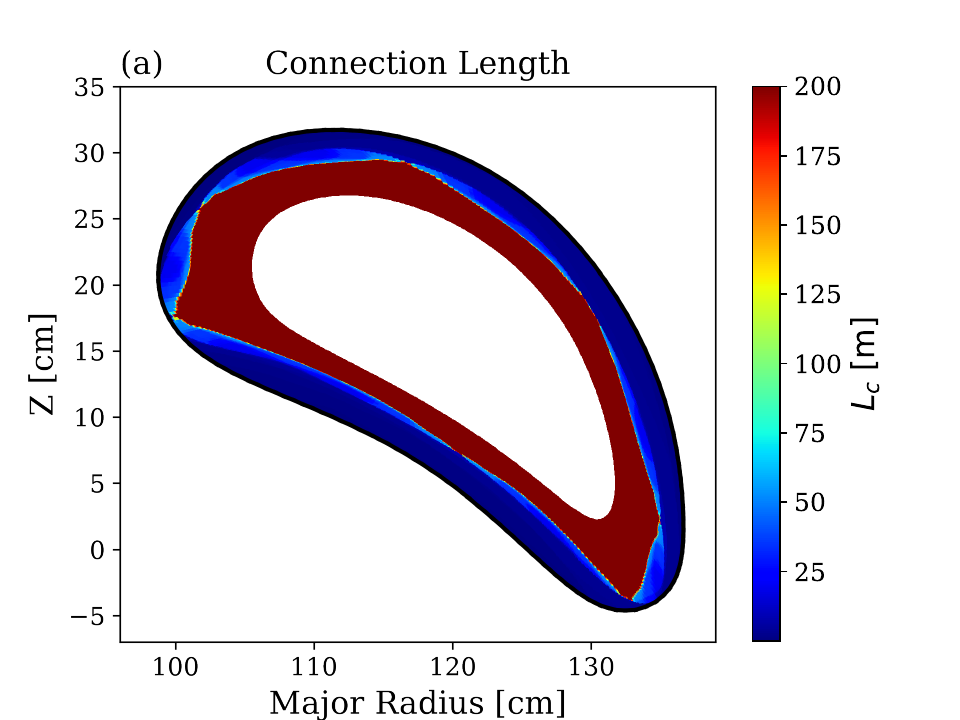}
\hfill
\includegraphics[width=0.49\textwidth]{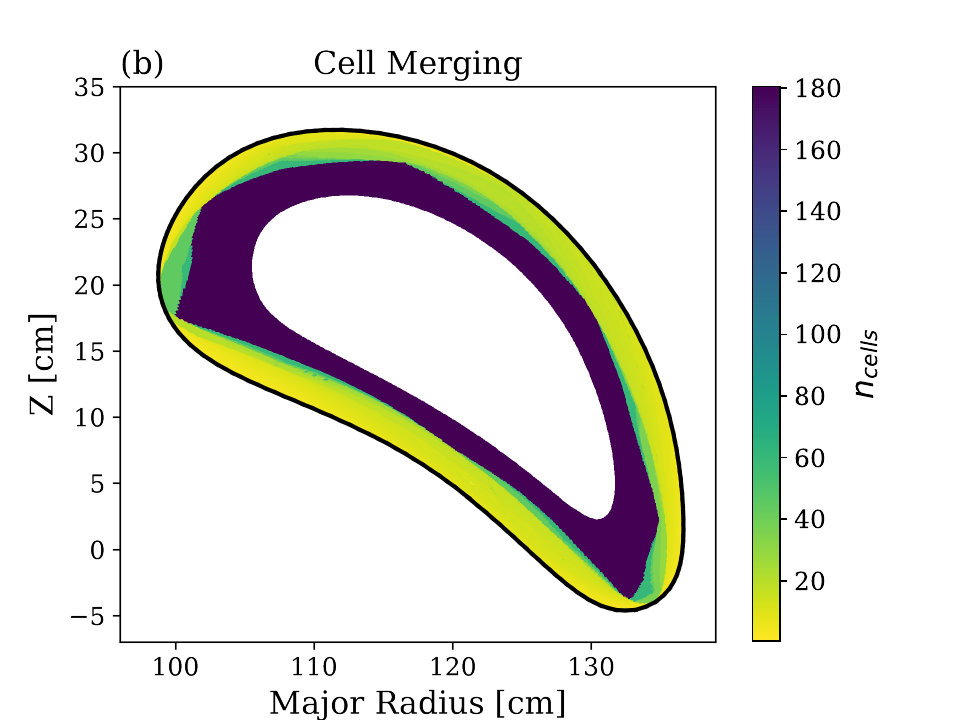}
\caption{R-Z plane at $\phi=22.5^\circ$ showing (a) connection length $L_c$, and (b) number of cells merged in field-aligned direction $n_{cells}$.}
\label{fig:Lc_merged_cells}
\end{figure}

The computational grid built with the FLARE code is well suited for storing and tracking MC particles. However, a memory problem can arise if the tracking of plasma and neutral quantities takes place on the same high fidelity grid. In order to reduce computational costs, an additional coarser so-called "physical mesh" is defined which scores and tracks volumetric plasma parameters via track length estimators weighted by MC particle residence time \cite{feng3DFluidModelling1999}. The physical mesh is built from the computational grid by merging grid cells in the field-aligned direction. Grid cells are scored based on their proximity to PFCs via field line tracing, and then contiguous field-aligned cells are partitioned, starting from the wall, to form the physical cells. This method ensures the physical mesh is fine near the PFCs but coarse in regions of closed flux surfaces. The argument for this scheme is that the fast parallel transport in the edge limits significant gradients from developing along field lines far from the PFCs. Therefore we may have a coarser field-aligned resolution in those regions. Note that neutral data is also stored on the physical mesh. For visualization, figure~\ref{fig:Lc_merged_cells} provides the connection length (wall-to-wall length of field lines) of HSX together with the implemented cell merging at the $\phi=22.5^\circ$ R-Z plane. To reduce the already high computational requirements needed for this study, we take the entire toroidal domain as our maximum physical cell length. This is appropriate as the maximum cell merging will only occur in the core of HSX. The extended neutral grid can also merge cells in the poloidal and toroidal directions, which further influences the simulation accuracy; however, this is not analyzed further in this study.

Despite Richardson extrapolation being generalizable to non-uniform and unstructured grids \cite{celikNumericalExperimentsApplication1997, roachePerspectiveMethodUniform1994}, it is not strictly valid in EMC3-EIRENE as geometric similarity of the grids is not preserved across refinements in the presence of cell merging. In other words, constant refinement cannot be guaranteed locally, and thus the firm basis of structured grid refinement is not applicable. The preferred method would be to use Richardson Extrapolation on grids with no cell merging, and compare with a solution in which cell merging is activated. However, if this were easily achievable, we would not merge grid cells at all.

To elucidate this point, we must first explain Richardson Extrapolation. Richardson Extrapolation relies on the assumption that the solution is smooth and differentiable and that the error lies within the asymptotic range $\epsilon_d \propto h^{p_d}$ where $h$ is a measure of grid spacing and $p_d$ is the order of convergence of the discretization error. We expect the discretization error to be second order for EMC3-EIRENE ($p_d=2$) \cite{ghoosNumericalAccuracyConvergence2020, ghoosAccuracyConvergenceIteratively2018}, but this can be confirmed with solutions on three successively finer grids. Using a refinement factor of $R=2$, the apparent order of convergence is determined as
\begin{equation}
    p_d = \frac{\log{ \left| {\frac{f_{h} - f_{2h}}{f_{0.5h} - f_{h}}} \right|}}{\log{(2)}}
    \label{eq:order_of_convergence}
\end{equation}
where $f_{0.5h}$, $f_h$, and $f_{2h}$ are, respectively, solutions on the high, medium, and low resolution grids. The absolute value in equation (\ref{eq:order_of_convergence}) ensures oscillatory convergence ($(f_h-f_{2h})/(f_{0.5h}-f_h)<0$) does not break this metric \cite{celikNumericalExperimentsApplication1997}. However, oscillatory convergence indicates we are not in the asymptotic regime and error estimates are unreliable \cite{colemanProblemOscillatoryBehavior2001, celikLimitationsRichardsonExtrapolation2005}. Therefore, the percentage occurrence of oscillatory convergence will also be reported \cite{ProcedureEstimationReporting2008}. The discretization error is calculated with an equation equivalent to (\ref{eq:bias_richardson}), and a safety factor of 1.25 is applied to the estimation to take into account inaccuracies in the estimate of $p_d$ \cite{boeyaertNumericalErrorAnalysis2022, roacheQuantificationUncertaintyComputational1997}.

Despite HSX being a relatively small device, evaluating three simulations on unmerged grids separated by the refinement factor $R=2$, while keeping all grids within the asymptotic regime, resulted in an impractically high computational cost. Due these complications, for this study, we analyze only the radial and poloidal discretization error. We only make refinements in the radial and poloidal directions $R_{(r,\theta)}$, while keeping the field-aligned discretization constant. This ensures structured refinement.

It is noted that additional discretization errors appear in EMC3-EIRENE. Transport across toroidal zones (often called "mapping interfaces") can introduce artificial numerical diffusion due to a geometric interpolation error on sheared grid lines. However, \cite{kobayshEMC3-EIRENEaccuracyconvergence} demonstrated that with a sufficiently fine grid, this numerical diffusion is upward of an order of magnitude smaller than the physical cross-field diffusion simulated. Consequently, this error becomes negligible in relation to the simulated plasma transport, and is ignored in this study. Grid misalignment at mapping interfaces and the radial boundary of the field-aligned grid also act as sinks, as EMC3 won't track MC particles which leave the field-aligned grid. This so-called "mapping loss" takes the form of loss terms in the simulation's particle and power balance, which must be monitored by users. The choice of using only one toroidal zone (section \ref{sec:simulation_setup}) further mitigates these effects. A complete analysis of the discretization error for EMC3-EIRENE simulations in stellarators is left for future work.






\section{Iteration procedure and data acquisition}
\label{sec:Simulation_approach}

For the iteration procedure we divide the simulation into two key stages: an initial transient stage and a post-processing stage where we average over plasma and neutral quantities to improve the statistics \cite{ghoosNumericalAccuracyConvergence2020}.

After initialization of the code, transient iterations are run, where quantities are changing towards a statistically stationary solution. The simulation solution in the transient regime is disregarded. Multiple methods were used to evaluate whether the solution was steady. First, the volume averaged relative change between the solution, iteration-to-iteration, (equation (\ref{eq:relative_change})) is small (typically $\leq 2\%$) \cite{baderSimulationsEdgeConfigurations2013}. Second, pressure-weighted average downstream plasma quantities ($\langle n_{i,d} \rangle$, $\langle T_{e,d} \rangle$, and $\langle T_{i,d} \rangle$) are visually inspected across iterations to ensure that no transient behavior remains in the solution. It is noted that upstream refers to the inner boundary of the field-aligned mesh, just inside the last closed flux surface (LCFS), and downstream refers to the region in front of the PFCs \cite{fengTransportIslandDivertors2002}. These average values are calculated across the entire computational domain. Third, volumetric data fields are observed between iterations to ensure no regions of substantial local change remain. Finally, the power balance of the simulation is tracked to ensure that a significant amount of power is not lost to grid irregularities (subsection \ref{subsec:discretization_error}). For this study, we are satisfied with $\leq 1\%$ power loss. The transient phase, for these HSX operating conditions, was typically found to last 20 to 30 iterations before the stationary regime was reached.

After the solution has reached a statistically stationary regime, we continue to iterate the simulation, where we calculate the post-processing average after every iteration. Finally, after $I$ number of iterations of post-process averaging, we load the averaged results and take that as our final solution. A similar strategy has previously been implemented in the SOLPS-ITER code package \cite{baelmansEfficientCodeSimulation2017, baelmansErrorAssessmentCode2017}. This post-processing method is a key strategy in error analysis, because it lets us reduce the statistical error, thus allowing a better measurement of deterministic errors. This is important as deterministic errors are estimated by comparing different simulations with an altered numerical parameter, and assuming the measured error is the dominant error between the solutions. Therefore, the order of error analysis here is similar to that used in \cite{boeyaertNumericalErrorAnalysis2022}, where we first analyze the statistical error $\epsilon_s$, then reduce it and measure the deterministic errors ($\epsilon_b$, $\epsilon_t$, $\epsilon_d$).

It is noted that all simulations are performed across 64 CPU cores. This is because varying the number of cores parallelized across introduced small deterministic offsets in the mean value of the simulation solution, given that proper adjustments were made to maintain the same total simulated MC particle count ($P=P_{input}\times n_{cores}$). This offset would interfere with error measurements and is therefore avoided.

\begin{figure}
\centering{\includegraphics[width = 0.7\textwidth]{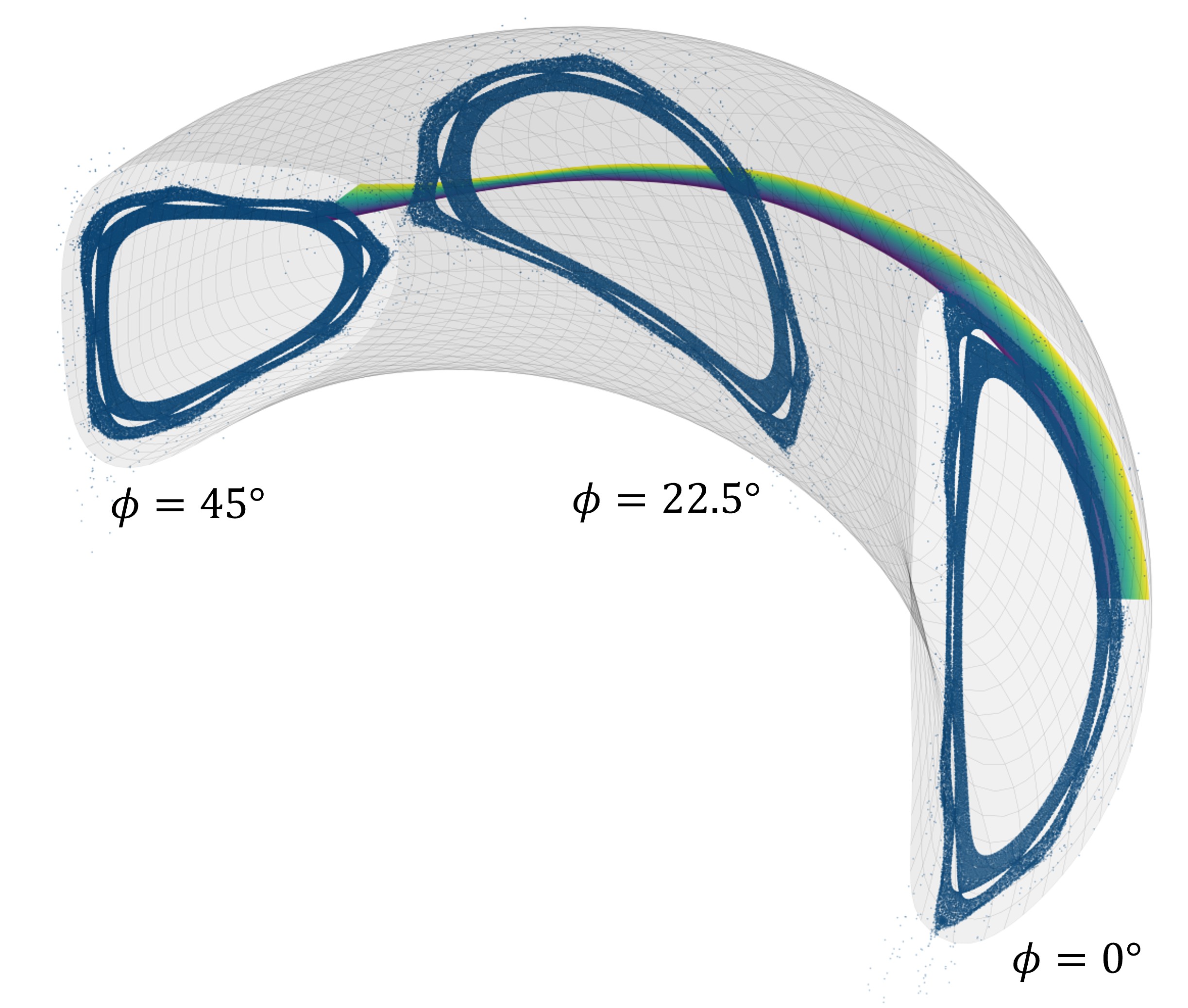}}
\caption{3D trajectory of field lines selected for plasma and neutral profile analysis. Poincar\'e plots at $\phi=0^\circ$, $22.5^\circ$, and $45^\circ$ are provided in blue for clarity. Note that the field line coloring is used purely for better visualization.}
\label{fig:field_lines}
\end{figure}

Errors are examined for volumetric quantities along both open and closed field lines. The volumetric quantities of interest include the plasma parameters ($n_i$, $M$, $T_e$, $T_i$) alongside neutral properties such as the atomic density $n_H$, atomic temperature $T_H$, and total neutral pressure $p_{tot,n}$. These are specifically chosen as they are fundamental in the characterization of plasma, both experimentally and computationally \cite{boeyaertImprovedNeutralExhaust2025}. 50 Field lines are traced in the forward direction (counter-clockwise) starting at varying locations along the outboard midplane at $\phi=0^\circ$. The outboard midplane is specifically chosen here as it crosses the LCFS, allowing for analysis of errors in a region of large gradients and therefore, large numerical errors. The field lines are shown in figure \ref{fig:field_lines}, as they travel from $\phi=0^\circ$ to $\phi=45^\circ$. This method is advantageous for stellarators, as field lines in different topological regions of the plasma edge can be analyzed individually. This also enables a general error description of all traced field lines. The error along the field lines is quantified as the L1-norm over field values $f_n$ and is written as
\begin{equation}
    \epsilon_{L1} = \frac{\sum_{n=1}^{N}|f_n - f_{n,ref}|}{\sum_{n=1}^{N}|f_{n,ref}|}
    \label{eq:L1-norm}
\end{equation}
where the reference solution $f_{n,ref}$ is defined as the most numerically accurate result available that is relevant to the error contribution under investigation. The reference value will vary throughout the study as we consider different error contributions. This approach follows the methodology established in \cite{ghoosNumericalAccuracyConvergence2020}, with necessary modifications for transitioning from tokamaks to stellarators.

In further scope of this paper, typically the asymptotic scaling of the error ($\epsilon_s$, $\epsilon_b$, and $\epsilon_t$) will be calculated by choosing a high fidelity simulation as reference and observing how changes in specific numerical parameters influence $\epsilon_{L1}$. Once the scaling of the error is observed, for the deterministic errors ($\epsilon_b$ and $\epsilon_t$), we then use Richardson-like extrapolation to obtain a reference solution, and calculate our final error estimation $\epsilon_{L1}$ with respect to that reference solution. This method is not suitable for the discretization error, and will be discussed and modified in subsection~\ref{subsec:results_discretization_error}.

\section{Error analysis in examined HSX case}
\label{sec:results}

\subsection{Statistical error}
\label{subsec:results_statistical_error}

The statistical error is quantified by first measuring the correlation time $\tau$ (equation (\ref{eq:auto_correlation})) independently for each volumetric quantity on every field line. The estimates of $\tau$ are calculated with the average value along a field line, which was stored after every iteration of the code. Data was stored across $1000$ iterations in five independent simulations following $P=10^6$ MC particles per iteration (for all modules), with parallel and perpendicular ENERGY time steps of $\Delta t_{E\parallel}=5\times10^{-8}$ s and $\Delta t_{E\perp}=2\times10^{-7}$ s, respectively, and a STREAMING time step of $\Delta t_S=5\times10^{-8}$ s (default time steps). $80\%$ of the simulations exhibited correlation times significantly higher on closed field lines compared to open field lines. To more rigorously observe this phenomenon, the simulation exhibiting the highest correlation times was isolated for an extended sampling of $5000$ iterations. The following estimates are obtained from this simulation and represent the largest $\tau$ values found, given only a finite number of iterations.

\begin{figure}
  \centering
  \includegraphics[width=0.48\textwidth]{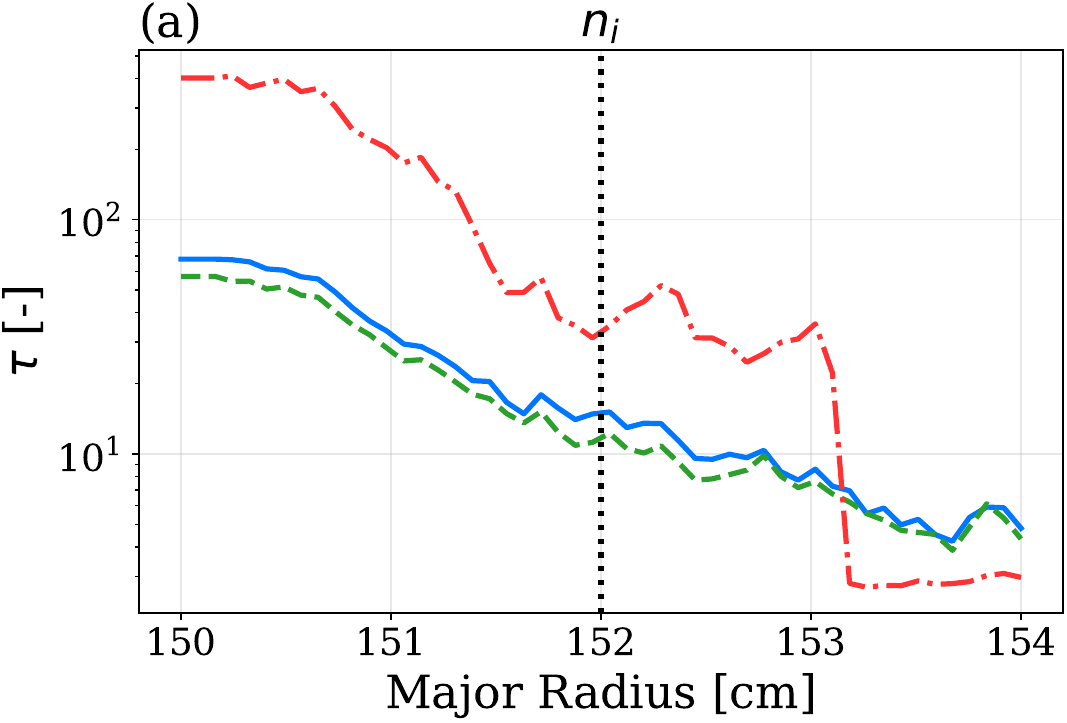}
  \hfill
  \includegraphics[width=0.48\textwidth]{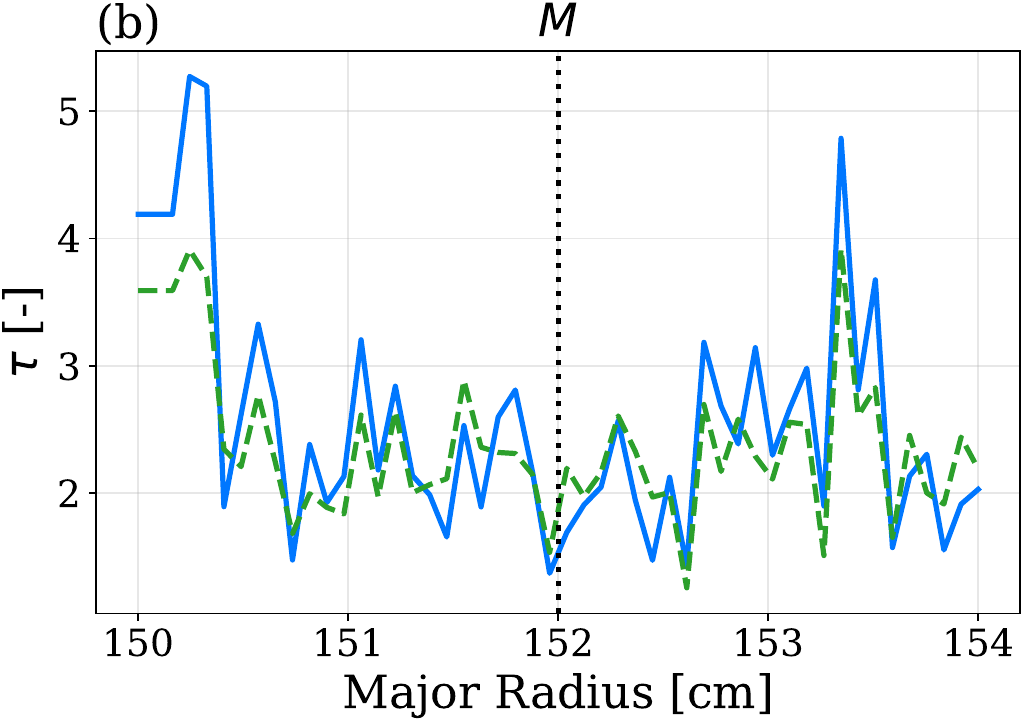}
  \vspace{0.3cm}
  \includegraphics[width=0.48\textwidth]{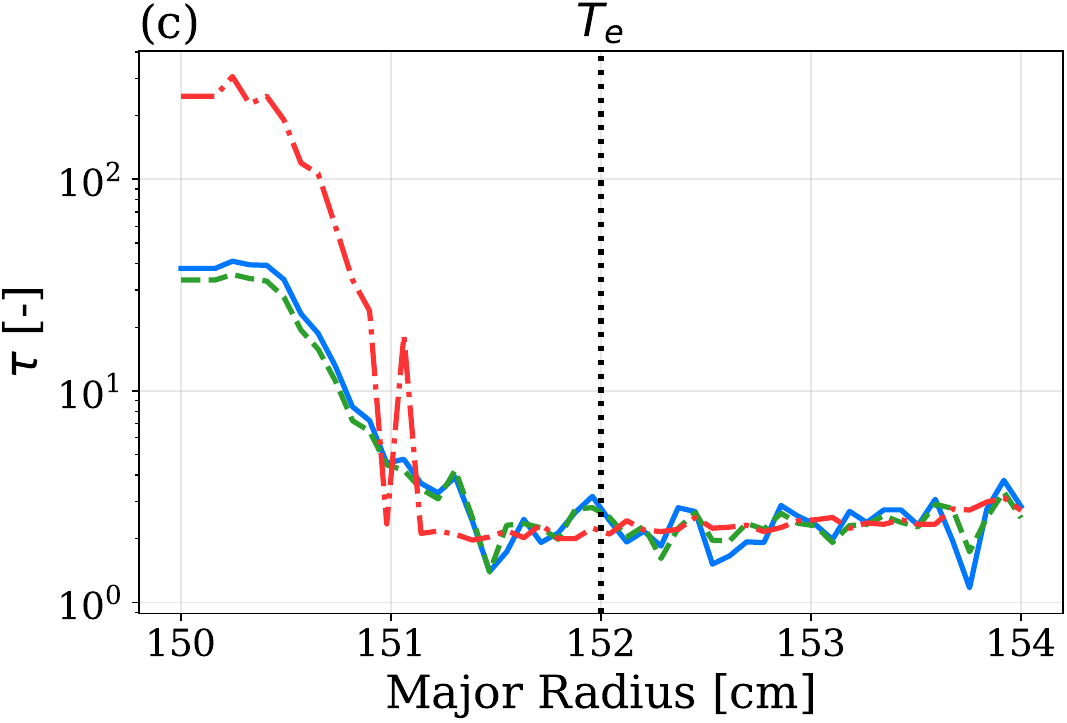}
  \hfill
  \includegraphics[width=0.48\textwidth]{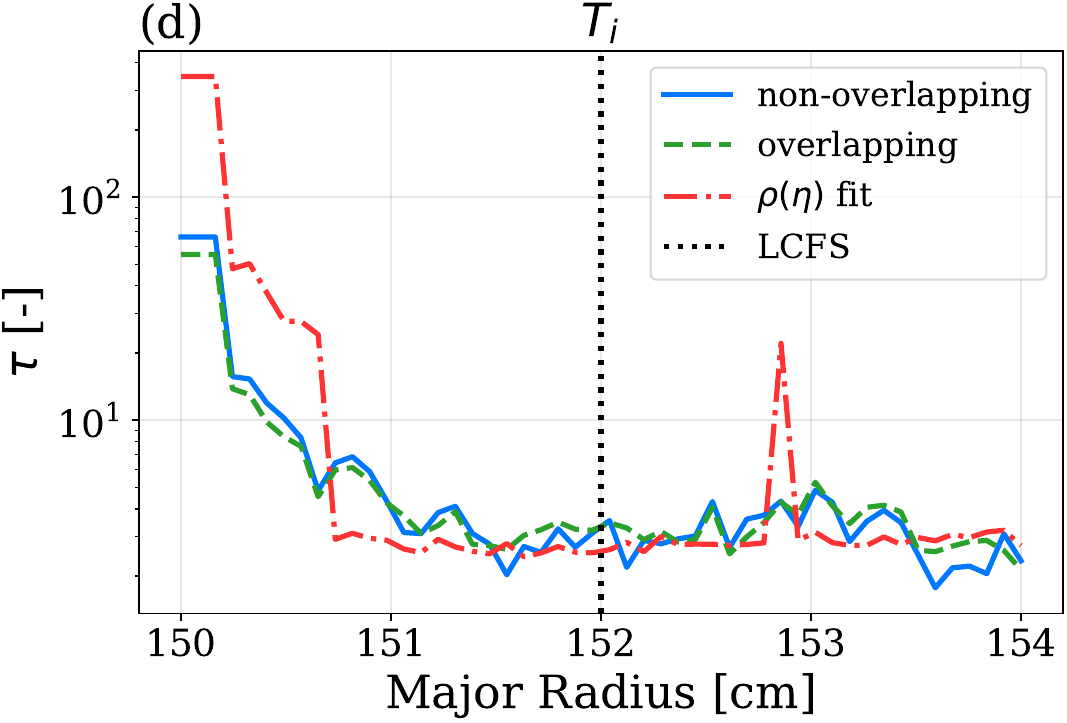}
  \caption{Radial profiles of the correlation time $\tau$ for (a) $n_i$, (b) $M$, (c) $T_e$, and (d) $T_i$ for the non-overlapping and overlapping batching methods and the autocorrelation method with a $\rho(\eta)$ exponential decaying fit. The x-axis represents the major radius of the field line traces at $\phi=0^\circ$. The approximate location of the LCFS is also shown.}
  \label{fig:correlation_profile}
\end{figure}

The radial profiles of $\tau$ for the plasma variables are provided in figure~\ref{fig:correlation_profile}. Radial profiles of neutral data are excluded due to a uniform and low estimate of $\tau$ throughout the domain. As can be seen in figure~\ref{fig:correlation_profile}, $\tau$ decreases radially outward for $n_i$, $T_e$, and $T_i$, indicating that the simulation evolves faster in the SOL than in the core region. This behavior may be attributed to imposing the upstream density and power boundary conditions at the innermost boundary of the simulation, which subsequently increases correlation there. The non-overlapping and overlapping batching methods and the autocorrelation method approximately produce $\tau$ estimates of the same order of magnitude in the SOL. However, the batching method incorrectly underestimates $\tau$ in the core region. This is the result of a large $\tau$ relative to the batch size $N$ (i.e. batches are still correlated at large $N$), which was determined by methods outlined in \cite{ghoosNumericalErrorEstimation2017}. Despite its limitations, the batching method did successfully estimate low $\tau$ values for $M$, $n_H$, $T_H$, and $p_{tot,n}$. With these low estimates, $\rho(\eta)$ is small and thus more influenced by statistical noise. Therefore, it was impossible to use the autocorrelation method to estimate $\tau$ for these variables. Table~\ref{tab:correlation_estimate} tabulates the largest $\tau$ value across all field lines for each method. It is seen that the neutral variables have a much smaller correlation time than the plasma variables. This is the result of the EIRENE code computing uncorrelated particle trajectories each iteration, meaning only the plasma background creates correlation. Similar results have been observed for neutral quantities in SOLPS-ITER \cite{ghoosNumericalErrorEstimation2017}.

\begin{table}
\caption{Largest $\tau$ estimate across all methods and field lines. '--' corresponds to estimates not possible due to noise dominating $\rho(\eta)$.}
\centering
\renewcommand{\arraystretch}{1.3}
\begin{tabular*}{\textwidth}{@{\extracolsep{\fill}}lccccccc}
\hline
$\tau$ estimation method & $n_i$ & $M$ & $T_e$ & $T_i$ & $n_H$ & $T_H$ & $p_{tot,n}$\\[6pt]
\hline
Batching: non-overlapping & 68 & 6 & 42 & 67 & 2 & 2 & 3\\
Batching: overlapping & 58 & 4 & 36 & 56 & 2 & 2 & 2\\
Autocorrelation: $\rho(\eta)$ fit & 412 & -- & 307 & 348 & -- & -- & --\\[6pt]
\hline
\end{tabular*}
\label{tab:correlation_estimate}
\end{table}

It is stressed that the estimates of $\tau$ are case specific, and different values will be found under different conditions. For the rest of this paper, the most conservative estimate of $\tau$ from table~\ref{tab:correlation_estimate} for each volumetric quantity is used to estimate the global statistical error. Due to the limited iterations available, this could be an underestimation in the core field lines, however, it overestimates $\tau$ for most open field lines and thus we are satisfied that it offers an upper bound of the statistical error for our quantities. 

To further characterize the statistical error, simulations were examined with varying MC particles counts, while keeping all other numerical parameters constant ($\Delta t_{E\parallel}= \Delta t_S=5\times10^{-8}$ s,  $\Delta t_{E\perp}=2\times10^{-7}$ s). A reference solution was used with $P=10^7$ MC particles (for all modules), which was averaged across $I=500$ iterations. Simulations were then performed, while varying the MC particle count of each module individually (respectively, $P_E$, $P_S$, and $P_N$ for the ENERGY, STREAMING, and NEUTRAL modules), while keeping the MC particle count of the other two modules not under investigation at the reference value. For example, to analyze the statistical error in $T_e$, we vary $P_E$, while keeping $P_S=P_N=10^7$. These simulations were also averaged across $I=500$ iterations.

It is noted that, for EMC3-EIRENE, the number of MC particles specified in the input file by the user is not directly equal to the number of tracked MC particles of the simulation. Rather, the user specifies a maximum weight for each particle, given the imposed source strength. Throughout the simulation, MC particles are dynamically split to combat downstream statistical noise resulting from volumetric sinks in front of the PFCs \cite{fengRecentImprovementsEMC3Eirene2014}. This splitting procedure usually leads to the tracked MC particle count being slightly higher than that imposed in the input file. Although an important procedure, the difference in MC particle counts is small, and therefore this effect will be ignored in this analysis. 

\begin{figure}
\centering{\includegraphics[width = 0.7\textwidth]{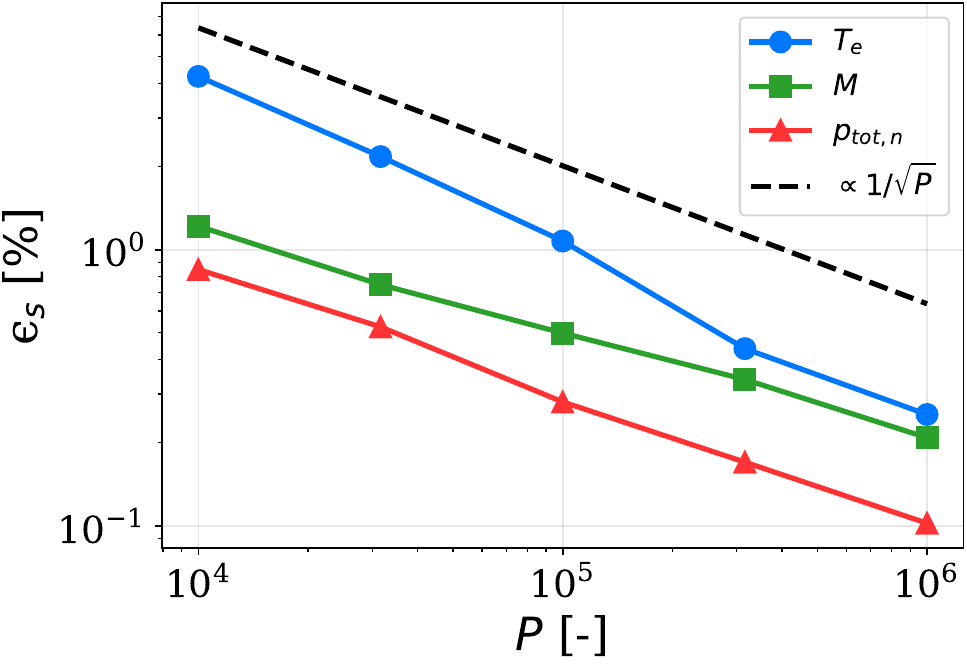}}
\caption{Statistical error $\epsilon_s$ in the L1-norm estimate (\ref{eq:L1-norm}) for $T_e$, $M$, and $p_{tot,n}$ for changes in $P_E$, $P_S$, and $P_N$, respectively. Note the MC particle count of each module: ENERGY $P_E$, STREAMING $P_S$, and NEUTRAL $P_N$. The L1-norm is taken with respect to a high fidelity reference solution with $P_E=P_S=P_N=10^7$. All simulations are averaged across $I=500$ iterations.}
\label{fig:stat_error_scaling}
\end{figure}

The asymptotic scaling of the statistical error is presented in figure \ref{fig:stat_error_scaling} for $T_e$, $M$, and $p_{tot,n}$ as we vary the MC particle count in the ENERGY, STREAMING, and NEUTRAL modules, respectively. These values represent the statistical uncertainty in our L1 error estimator (\ref{eq:L1-norm}) across all field lines with respect to the higher fidelity reference simulation whose residual statistical error is assumed negligible. An additional line $\propto 1/\sqrt{P}$ is provided to display the reduction rate of $\epsilon_s$, which is consistent with equation (\ref{eq:stat_error}). The other variables tracked are omitted for clarity, but follow the same trend as those presented. An ordinary least-squares method was used to fit the statistical error, preserving the $1/\sqrt{P}$ scaling, to estimate the value of $A_s$ in equation (\ref{eq:stat_error}) for each variable. These values are provided in Appendix~\ref{sec:appendix_A}.

Furthermore, estimates of the statistical error for varying MC particle counts, given no post-process averaging ($I=1$), are provided in table~\ref{tab:stat_error}. These estimates are therefore larger than those in figure~\ref{fig:stat_error_scaling} by a factor of $\sqrt{500} \approx22.4$. These values are particularly interesting because post-process averaging is not yet widely used in the EMC3-EIRENE community. As a result, we may gain additional insight from these estimates when compared to the iteration-to-iteration volume averaged relative change (\ref{eq:relative_change}) of our plasma quantities on the chosen field lines. From this table two conclusions are clear. First, the electron and ion temperatures are the most sensitive to numerical noise in the simulation, especially in low MC particle simulations. This may change in front of the strike lines or as simulations reach a detached regime where temperatures can be low ($<10$ eV). Second, the commonly used volume averaged relative change metric is insufficient as a standalone estimator for determining the statistical error in simulations as it systematically underestimates the error on the simulation solution for plasma quantities. The measure fails because it tracks only the iteration-to-iteration change between potentially highly correlated iterations rather than the deviation from the true expected solution. This is seen as the only relative change measure within the correct order of magnitude as the statistical error is for $M$, which was measured to have a relatively small correlation time. This point in turn supports the need for post-process averaging, as it is a useful procedure in further reducing the statistical error in a steady solution, even if the relative change is small. Although not shown, it is reported that the statistical error on volumetric quantities was roughly constant across the radial scan of field lines. This is interpreted for plasma quantities as higher variance but lower correlation downstream balancing with lower variance but higher correlation upstream, and for neutral quantities as equal correlation and variance throughout the chosen field lines. 

\begin{table}
\caption{Statistical error $\epsilon_s$~[\%] for MC particle counts of each module: ENERGY $P_E$, STREAMING $P_S$, and NEUTRAL $P_N$, without post-process averaging ($I=1$). '--' corresponds to estimates in which the statistical error is negligible on a quantity. In addition, the volume averaged relative change [\%] (\ref{eq:relative_change}), averaged across $500$ steady iterations, is provided for plasma quantities.}
\centering
\renewcommand{\arraystretch}{1.3}
\begin{tabular*}{\textwidth}{@{\extracolsep{\fill}}lccccccccccc}
\hline
$P$ & $n_i$ & $M$ & $T_e$ & $T_i$ & $n_H$ & $T_H$ & $p_{tot,n}$ & $\Delta n_i$ & $\Delta M$ & $\Delta T_e$ & $\Delta T_i$ \\
\hline

$P_E=10^4$ & -- & -- & $94.9$ & $129.2$ & -- & -- & -- & $0.46$ & $5.13$ & $5.70$ & $7.10$ \\
$P_E=10^5$ & -- & -- & $24.0$ & $22.8$ & -- & -- & -- & $0.19$& $2.8$ & $2.02$ & $2.51$ \\
$P_E=10^6$ & -- & -- & $5.7$ & $6.5$ & -- & -- & -- & $0.12$ & $2.35$ & $0.63$ & $0.79$\\[6pt]
\hline

$P_S=10^4$ & $30.8$ & $27.2$ & -- & -- & -- & -- & -- & $2.00$ & $33.55$ & $0.42$ & $0.63$\\
$P_S=10^5$ & $18.4$ & $11.2$ & -- & -- & -- & -- & -- & $0.84$ & $8.14$ & $5.51$ & $4.22$ \\
$P_S=10^6$ & $7.8$ & $4.7$ & -- & -- & -- & -- & -- & $0.33$ & $3.5$ & $2.37$ & $1.84$\\[6pt]
\hline

$P_N=10^4$ & -- & -- & -- & -- & $24.1$ & $26.8$ & $19.0$ & $0.14$ & $1.26$ & $0.91$ & $0.75$\\
$P_N=10^5$ & -- & -- & -- & -- & $5.9$ & $7.6$ & $6.3$ & $0.12$ & $1.22$ & $0.88$ & $0.72$ \\
$P_N=10^6$ & -- & -- & -- & -- & $1.6$ & $2.7$ & $2.3$ & $0.12$ & $1.21$ & $0.87$ & $0.72$ \\[6pt]
\hline
\end{tabular*}
\label{tab:stat_error}
\end{table}

Regarding the dependence of the statistical error on the grid resolution $h$, it was observed that the variance $\sigma_1^2$ of volumetric parameters with low correlation ($M$, $n_H$, $T_H$, and $p_{tot,n}$) scaled with grid refinement $R_{(r,\theta)}$ similarly to that reported in \cite{ghoosNumericalAccuracyConvergence2020}. This scaling is recorded in Appendix~\ref{sec:appendix_A}. For the variables with high correlation ($n_i$, $T_e$, and $T_i$), the scaling of $\sigma_1^2$ has not been verified as this would take an extensive sample size to accurately capture $\sigma_1^2$. For these variables, to be conservative, we take the largest observed scaling among $M$, $n_H$, $T_H$, and $p_{tot,n}$. This also indicates the bias's dependence on grid resolution, which is similarly reported in Appendix~\ref{sec:appendix_A}. For further analysis in this study, $\epsilon_s$ is reduced using both the MC particle count and iterations of post-process averaging, which will be discussed alongside subsequent error measurements. 

\subsection{Bias}
\label{subsec:results_bias}

Richardson-like extrapolation relies on the assumption that the bias is the asymptotic regime. However, background statistical noise often obscures the verification of the $\epsilon_b \propto 1/P$ scaling. To isolate the bias, we adopted the decoupling approach established in \cite{ghoosNumericalAccuracyConvergence2020}.

First, a fully coupled high fidelity reference simulation is performed with $P_E=P_S=P_N=1 \times 10^7$ MC particles, $\Delta t_{E,\parallel}=\Delta t_S=5 \times 10^{-8}$ s, $\Delta t_{E,\perp}=2 \times 10^{-7}$ s, and $I=500$. We then run several decoupled simulations in which the MC particle count of a single module is varied to ensure only the quantities calculated in the module are updated while the plasma and neutral profiles calculated in the other modules remain the same as the reference case. For example, to analyze the bias in $T_e$, we first initialize a simulation, then upload the STREAMING and NEUTRAL module results from the high fidelity reference simulation, and vary $P_E$ while keeping all other numerical parameters constant. Then transient iterations towards a steady solution and post-processing averaging ($I=1000$) are performed where we only iterate the ENERGY module. By evaluating the L1-norm (\ref{eq:L1-norm}) with respect to the high fidelity reference simulation, we characterize the bias's dependence on $P_E$. Note that this strategy is only used to observe the scaling of the bias, and we return to fully coupled EMC3-EIRENE simulations to estimate the magnitude, as well as cross module dependence.

\begin{figure}
\centering{\includegraphics[width = 0.7\textwidth]{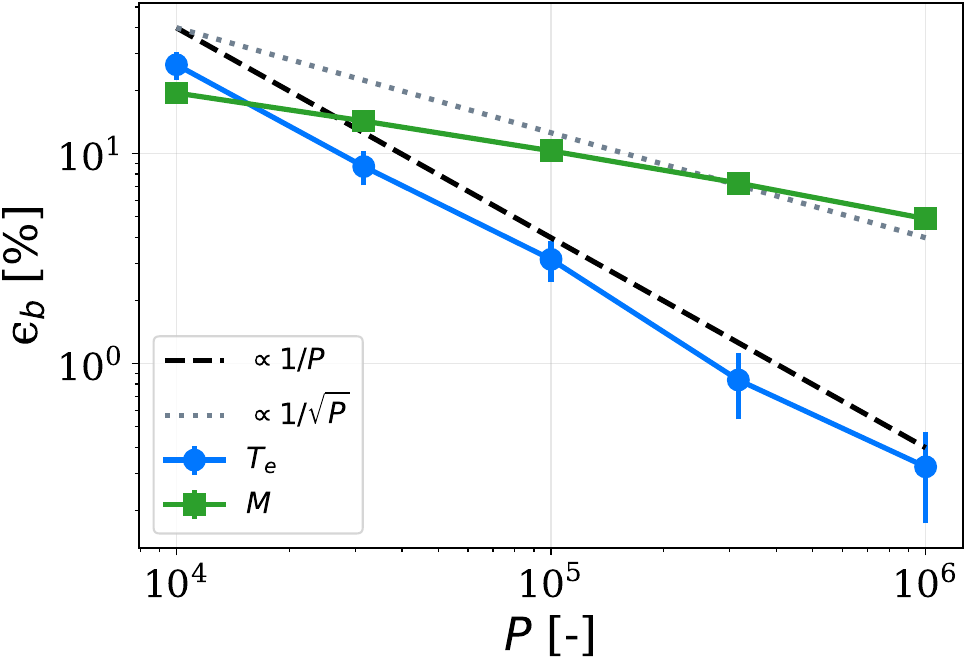}}
\caption{Bias $\epsilon_b$ scaling in $T_e$ and $M$ for changes in $P_E$ and $P_S$, respectively. Note the MC particle count of each EMC3 module: ENERGY $P_E$ and STREAMING $P_S$.}
\label{fig:bias_scaling}
\end{figure}

Figure \ref{fig:bias_scaling} presents the resulting bias for $T_e$ and $M$ as we vary the MC particle count in the ENERGY and STREAMING modules, respectively. These values represent the global L1 error across all field lines. Two reference lines are provided: $\propto 1/P$ to display the theoretical reduction rate and $\propto 1/\sqrt{P}$ for interpretation. Although the bias for $T_e$ closely follows the theoretical scaling, the scaling in $M$ is closer to $\propto 1\sqrt{P}$. This is attributed to the remaining statistical error dominating estimates. However, even if remaining noise prevents the theoretical reduction rate from distinctly appearing, we can still estimate the magnitude of the bias with Richardson-like extrapolation \cite{ghoosNumericalAccuracyConvergence2020}. Due to the implementation of the NEUTRAL module in the code, we cannot isolate it as done with the plasma modules. Therefore, we assume the theoretical scaling for $n_H$, $T_H$, and $p_{tot,n}$.


Using four fully coupled simulations with varying MC particle counts in the ENERGY module ($P_E=10^4$, $10^5$, $10^6$, and $10^7$) alongside fixed parameters ($P_S=P_N=10^7$, $\Delta t_{E,\parallel}=\Delta t_S=5 \times 10^{-8}$ s, $\Delta t_{E,\perp}=2 \times 10^{-7}$ s, and $I=500$), equation (\ref{eq:bias_richardson}) can be applied to estimate the bias across changes in $P_E$. All simulations are compared with the lowest fidelity simulation ($P_E=10^4$). This procedure is repeated for the STREAMING ($P_S$) and NEUTRAL ($P_N$) modules to isolate the bias resulting from each MC particle population. The resulting bias estimates are presented in table \ref{tab:bias_error}. These estimates represent the L1-norm of the bias where, at all field values, we assume the theoretical order and extrapolate the solution $f_{P\rightarrow \infty}$ as our reference solution. The convergence order may vary locally, especially in regions of sharp gradients or flow reversal \cite{AlexeevOnSomeFeaturesofRichardsonExtrapolation}. Therefore, the estimated values should be interpreted as an approximate global estimation rather than an exact point-wise estimation.

\begin{table}
\caption{Bias $\epsilon_b~[\%]$ for MC particle counts of each module in EMC3-EIRENE: ENERGY $P_E$, STREAMING $P_S$, and NEUTRAL $P_N$. Estimates marked with an asterisk ($^*$) denote cases where $\epsilon_b < \epsilon_s$, and thus carry high uncertainty.}
\centering
\renewcommand{\arraystretch}{1.3}
\begin{tabular*}{\textwidth}{@{\extracolsep{\fill}}lccccccc}
\hline
$P$ & $n_i$ & $M$ & $T_e$ & $T_i$ & $n_H$ & $T_H$ & $p_{tot,n}$\\
\hline

$P_E=10^4$ & $7.74$ & $12.04$ & $16$ & $25$ & $11.51$ & $10.99$ & $9.45$ \\
$P_E=10^5$ & $0.77$ & $1.20$ & $2$ & $2$ & $1.15$ & $1.10$ & $0.95$ \\
$P_E=10^6$ & $0.08$ & $0.14$ & $0.2^*$ & $0.2^*$ & $0.11$ & $0.13$ & $0.08$ \\
$P_E=10^7$ & $0.008^*$ & $0.01^*$ & $0.02^*$ & $0.02^*$ & $0.01^*$ & $0.02^*$ & $0.01^*$ \\[6pt]
\hline

$P_S=10^4$ & $2$ & $15$ & $1.2$ & $2.8$ & $10.65$ & $6.39$ & $10.92$ \\
$P_S=10^5$ & $0.2^*$ & $1.5$ & $0.1^*$ & $0.3$ & $1.07$ & $0.64$ & $1.09$ \\
$P_S=10^6$ & $0.01^*$ & $0.2$ & $0.01^*$ & $0.04^*$ & $0.10$ & $0.06$ & $0.09$ \\
$P_S=10^7$ & $0.001^*$ & $0.02^*$ & $0.001^*$ & $0.002^*$ & $0.01^*$ & $0.005^*$ & $0.01^*$ \\[6pt]
\hline

$P_N=10^4$ & $3.38$ & $2.47$ & $1.5$ & $3.4$ & $177.7$ & $108$ & $141.3$ \\
$P_N=10^5$ & $0.34$ & $0.28$ & $0.15$ & $0.3$ & $17.8$ & $10.8$ & $14.1$ \\
$P_N=10^6$ & $0.03$ & $0.01^*$ & $0.02^*$ & $0.03^*$ & $1.50$ & $0.98$ & $1.27$ \\
$P_N=10^7$ & $0.002^*$ & $0.001^*$ & $0.001^*$ & $0.003^*$ & $0.15$ & $0.10$ & $0.13$ \\[6pt]
\hline
\end{tabular*}
\label{tab:bias_error}
\end{table}





The following two observations can be made from table \ref{tab:bias_error}. First, the bias is distinctively largest for the neutral data when varying $P_N$. This provides quantitative support for the common practice in EMC3-EIRENE modeling of simulating more neutral MC particles than plasma MC particles. Second, the bias as a whole becomes negligible for these simulations at high MC particle counts ($P \ge 10^7$) as the statistical error will clearly dominate with $I=1$ (see table \ref{tab:stat_error} compared with table \ref{tab:bias_error}). Similar to \cite{ghoosNumericalAccuracyConvergence2020}, Richardson-like extrapolation was used to parameterize the bias, and the results are provided in Appendix \ref{sec:appendix_A}.

\subsection{Time integration error}
\label{subsec:results_time_integration_error}

\begin{figure}
\centering
\includegraphics[width=0.49\textwidth]{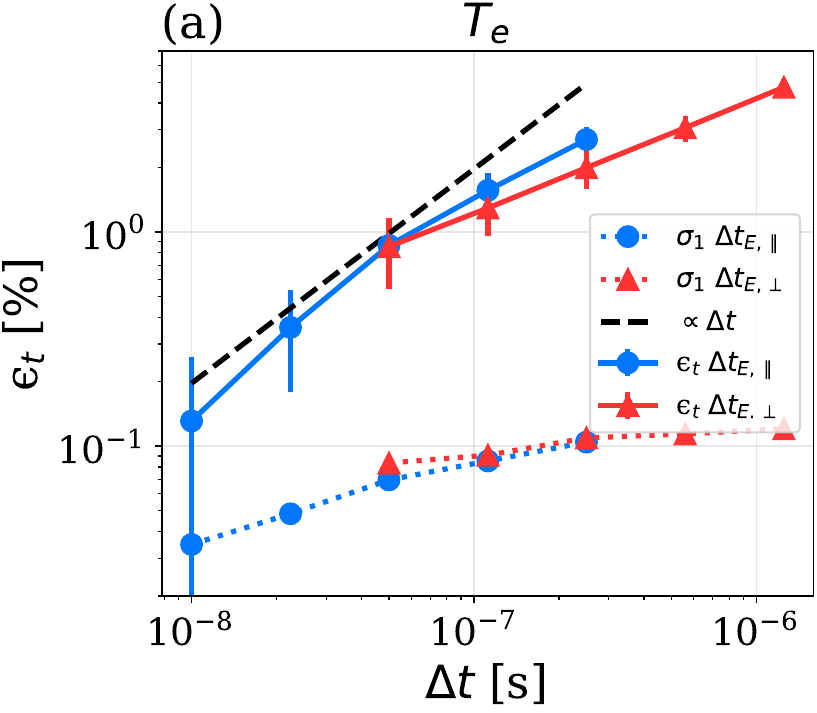}
\hfill
\includegraphics[width=0.49\textwidth]{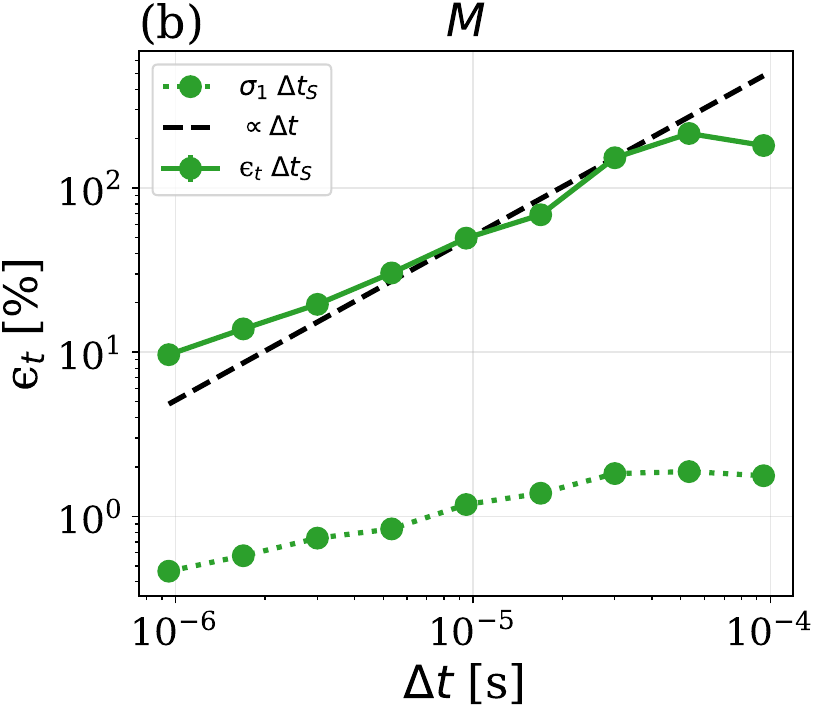}
\caption{Time integration error $\epsilon_t$ scaling of (a) $T_e$ and (b) $M$ for changes in (a) $\Delta t_{E\parallel}$, $\Delta t_{E\perp}$, and (b) $\Delta t_S$. The standard deviation of iterations $\sigma_1$ is also provided.}
\label{fig:time_integration_error_scaling}
\end{figure}

Similarly to the bias, the asymptotic regime of the time integration error is explored with decoupled simulations. Using the results of a fully coupled reference simulation ($\Delta t_{E,\parallel}=\Delta t_S = 1\times10^{-9}$ s, $\Delta t_{E,\perp}=5\times10^{-9}$ s, $P_E=P_S=P_N=5\times10^6$, and $I=200$), we run a series of decoupled simulations varying only one time step input. For these runs, only the ENERGY or STREAMING module is individually iterated, while the other modules are fixed. Figure \ref{fig:time_integration_error_scaling} presents the time integration error $\epsilon_t$ for $T_e$ and $M$ as we vary the different time steps. To clearly observe the scaling of the error in the STREAMING module, the measurements were performed at time steps two orders of magnitude greater than that used for the ENERGY module. The ENERGY module was unable to be analyzed at these high time steps ($\Delta t_{E,\parallel} \ge 5\times10^{-7}$ s) because EMC3-EIRENE would experience numerical instability and not converge. In addition to the time integration error, these plots present the standard deviation of iterations $\sigma_1$ of our estimate as a function of time step to characterize the statistical error's dependence on the time step.

From figure~\ref{fig:time_integration_error_scaling}, it is clear that the theoretical scaling is approximately true ($\epsilon_t \propto \Delta t_{E,\parallel},\Delta t_{E,\perp},\Delta t_S$) for $T_e$ and $M$. The time integration error in $T_e$ is particularly interesting as the overlap region between $\Delta t_{E,\parallel}$ and $\Delta t_{E,\perp}$ experiences error on the same order of magnitude, indicating that the common practice of choosing $\Delta t_{E,\perp}$ an order of magnitude larger than $\Delta t_{E,\parallel}$ is not always an error free assumption. This distinctly differs from the results of \cite{ghoosNumericalAccuracyConvergence2020}, and can be attributed to SOL cross-field transport playing a more prominent rule in stellarators versus tokamaks \cite{fengComparisonStellaratorTokamak2011} and the difference in diffusion and diffusivity values taken in these simulations compared to those taken in \cite{ghoosNumericalAccuracyConvergence2020}. For the error in $M$, we see that the asymptotic range is left at roughly $\Delta t_S \approx 10^{-4}$, and remaining statistical noise begin to influence our estimates at smaller time steps ($\Delta t_S \le 3 \times 10^{-6}$). It is also seen for both $T_e$ and $M$ that $\sigma_1$ shows a slight dependence on the time step in uncoupled simulations. However, in fully coupled simulations this dependence was observed to be negligible, similar to the results of \cite{ghoosNumericalAccuracyConvergence2020}, and thus is ignored in this analysis.

\begin{table}
\caption{Time integration error $\epsilon_t~[\%]$ for each EMC3 time step: parallel ENERGY time step $\Delta t_{E\parallel}$, perpendicular ENERGY time step $\Delta t_{E\perp}$, and the STREAMING time step $\Delta t_S$. Estimates marked with an asterisk ($^*$) denote cases where $\epsilon_t < \epsilon_s$, and thus carry high uncertainty.}
\centering
\renewcommand{\arraystretch}{1.3}
\begin{tabular*}{\textwidth}{@{\extracolsep{\fill}}lcccc}
\hline
$\Delta t$ & $n_i$ & $M$ & $T_e$ & $T_i$ \\
\hline

$\Delta t_{E\parallel} = 2.5 \times 10^{-7}$ & $1.8$ & $3.4$ & $3.2$ & $5.6$ \\
$\Delta t_{E\parallel} = 5 \times 10^{-8}$ & $0.4$ &$ 0.7$ & $0.6$ & $1.1$ \\
$\Delta t_{E\parallel} = 1 \times 10^{-8}$ & $0.04^*$ & $0.1^*$ & $0.1^*$ & $0.09^*$ \\[6pt]
\hline

$\Delta t_{E\perp} = 2.5 \times 10^{-7}$ & $0.3$ & $3.3$ & $1.8$ & $1.4$ \\
$\Delta t_{E\perp} = 5 \times 10^{-8}$ & $0.05^*$ & $0.7$ & $0.4^*$ & $0.3^*$ \\
$\Delta t_{E\perp} = 1 \times 10^{-8}$ & $0.05^*$ & $0.2$ & $0.1^*$ & $0.1^*$ \\[6pt]
\hline

$\Delta t_{S} = 2.5 \times 10^{-5}$ & $10.9$ & $78.6$ & $12.8$ & $9.2$ \\
$\Delta t_{S} = 5 \times 10^{-6}$ & $2.2$ & $15.7$ & $2.6$ & $1.8$ \\
$\Delta t_{S} = 1 \times 10^{-6}$ & $0.5$ & $3.5$ & $0.5$ & $0.4^*$ \\[6pt]
\hline
\end{tabular*}
\label{tab:time_integration_error}
\end{table}

Using fully coupled simulations with varying time steps ($P_E=P_S=P_N=5\times10^6$ and $I=100$), we can estimate the time integration error with respect to each time step. Similar to the bias, we compare simulation solutions assuming the theoretical order of convergence at all field points, and estimate the L1-norm with respect to the extrapolated reference solution $f_{\Delta t \rightarrow0}$. The results are presented in table \ref{tab:time_integration_error}. Firstly it is noted that deviations in neutral quantities from changes in $\Delta t$ are so small that $\epsilon_s > \epsilon_t$ despite averaging, and thus Richardson-like extrapolation is not valid. This simply indicates neutral quantities are roughly independent of $\Delta t$, and thus are excluded from table \ref{tab:time_integration_error}. This is not a surprising result as neutral quantities would only be affected through the plasma quantities, which themselves are seen to overall change very little. It is further seen in the table that the error resulting from $\Delta t_{E,\parallel}$ and $\Delta t_{E,\perp}$ is larger than the error resulting from $\Delta t_S$, when comparing time steps of similar magnitude. The conclusion is that $\Delta t_S$ can be chosen greater than $\Delta t_{E,\parallel}$ and $\Delta t_{E,\perp}$ to speed up simulation time without a loss to accuracy. It was also observed that the magnitude of the time integration error did not change significantly with grid resolution. The previous two points agree with the results in \cite{ghoosNumericalAccuracyConvergence2020}. Richardson-like extrapolation was used to parameterize the time integration error, and the results are provided in Appendix~\ref{sec:appendix_A}.

\subsection{Discretization error}
\label{subsec:results_discretization_error}

The discretization error along the field line profiles could not be analyzed with Richardson Extrapolation under only radial and poloidal grid refinements due to cross-coordinate coupling to the the field-aligned coordinate direction. To clarify this point, table \ref{tab:local_discretization_order} represents three separate $p_d$ measurements (equation (\ref{eq:order_of_convergence})) for grids with differing field-aligned resolution. We take the main grid utilized in this study as the medium resolution ($f_{h}$) grid, and apply radial and poloidal refinements of $R_{(r,\theta)}=2$ to obtain a low and a high resolution grid. We repeat this for two additional medium resolution grids where we slightly change the resolution of the field-aligned direction: in one grid we only increase the computational grid resolution and in the other we only increase the physical grid resolution. Each of the nine implemented grids is constructed separately to ensure field alignment. To prevent the previous error contributions from interfering with these measurements, the numerical parameters were chosen such that the net error $\epsilon_s+\epsilon_b+\epsilon_t <1\%$ for all grids. From table \ref{tab:local_discretization_order}, it is seen that the calculated orders differ significantly from the theoretical order ($p_d=2$). Furthermore, despite each measurement representing the same radial and poloidal resolutions, the $p_d$ values differ significantly across varying field-aligned resolutions. The high occurrence of oscillatory convergence also indicates this field-aligned dependence, and further justifies why Richardson Extrapolation cannot give suitable error estimates on this data. It is noted that this behavior distinctly differed from the bias and time integration error, where monotonic convergence was observed.

\begin{table}
\caption{Average local order of convergence $p_d$ across radial and poloidal grid refinements $R_{(r,\theta)}=2$ for grids with varying computational resolution ($r \times \theta \times \phi$), and maximum physical cell resolution in the field-aligned direction $h_{\text{physical,max}}$. The percent occurrence of oscillatory convergence is also provided.}
\centering
\renewcommand{\arraystretch}{1.3}
\begin{tabular*}{\textwidth}{@{\extracolsep{\fill}}lccccccc} \hline  & $n_i$ & $M$ & $T_e$ & $T_i$ & $n_H$ & $T_H$ & $p_{tot,n}$\\
\multicolumn{8}{l}{$f_{h}$ Grid: $150 \times 400 \times 180$ ; $h_{\text{physical,max}}=45^\circ$} \\
Apparent $p_d$               & $1.03$ & $0.47$ & $-0.25$ & $0.59$ & $0.14$ & $0.09$ & $0.001$ \\
Oscillatory Convergence $[\%]$      & $67.58$ & $60.37$ & $44.57$ & $30.65$ & $46.98$ & $54.59$ & $61.60$ \\[6pt]
\hline
\multicolumn{8}{l}{$f_{h}$ Grid: $150 \times 400 \times 216$ ; $h_{\text{physical,max}}=45^\circ$} \\
Apparent $p_d$               & $0.81$ & $1.02$ & $-0.11$ & $-0.06$ & $0.12$ & $0.32$ & $0.20$ \\
Oscillatory Convergence $[\%]$      & $54.24$ & $48.10$ & $28.60$ & $0.87$ & $50.41$ & $46.44$ & $62.16$ \\[6pt]
\hline
\multicolumn{8}{l}{$f_{h}$ Grid: $150 \times 400 \times 180$ ; $h_{\text{physical,max}}=22.5^\circ$} \\
Apparent $p_d$               & $-0.03$ & $-0.23$ & $-0.37$ & $-1.52$ & $0.86$ & $-0.50$ & $0.24$ \\
Oscillatory Convergence $[\%]$      & $68.02$ & $50.25$ & $20.11$ & $5.71$ & $55.20$ & $59.45$ & $56.52$ \\[6pt]
\hline
\end{tabular*}
\label{tab:local_discretization_order}
\end{table}

Cross-coordinate coupling of field data under independent coordinate refinements is well known in the computational fluid dynamics (CFD) community to produce unusable error estimates \cite{roachePerspectiveMethodUniform1994}. Furthermore, the presence of strong grid stretching by our cell merging scheme exacerbates this issue. However, solution functionals, which Richardson Extrapolation can also be applied on, should not experience these shortcomings \cite{roacheQuantificationUncertaintyComputational1997}. We therefore shift our analysis to pressure-weighted average downstream plasma quantities such as the downstream density $\langle n_{i,d} \rangle$ and downstream electron and ion temperatures ($\langle T_{e,d} \rangle$, and $\langle T_{i,d} \rangle$). It is noted that the convergence order on solution functionals scales the same as the local field data. The entire above analysis was repeated for these functionals and parameterization of the error on these quantities is provided in Appendix~\ref{sec:appendix_A}. The convergence order across radial and poloidal grid refinements was measured to be $p_d=0.07$, $0.92$, and $0.78$ for $\langle n_{i,d} \rangle$, $\langle T_{e,d} \rangle$, and $\langle T_{i,d} \rangle$, respectively. No oscillatory convergence was observed. Measurements of $p_d$ across varying field-aligned resolutions differed notably less than the field data and only deviated by a maximum of $0.19$, $0.33$, and $0.31$ for our respective plasma quantities. This is attributed to remaining cross-coordinate dependencies and the effect of cell merging. With these measured $p_d$ values, we can estimate the error on our solution functionals, which are presented in table~\ref{tab:functional_discretization_estimates}. These estimates are normalized by the extrapolated continuum solution. The estimates of $\epsilon_d$ for $\langle n_{i,d} \rangle$ are very conservative, resulting from the measured $p_d=0.07$, while the measures for $\langle T_{e,d} \rangle$ and $\langle T_{i,d} \rangle$ are more reliable and align closer with estimates using $p_d=2$. Overall it is reported that $\epsilon_d$ is larger in magnitude than $\epsilon_s$, $\epsilon_b$, and $\epsilon_t$ for typical simulation parameters despite only considering error from the radial and poloidal discretization. This highlights the need for adequately refined computational grids in EMC3-EIRENE simulations.

\begin{table}
\caption{Discretization error $\epsilon_d~[\%]$ of pressure-weighted average downstream plasma quantities across radial and poloidal grid refinements $R_{(r,\theta)}=2$ (Grid: $r \times \theta \times \phi$). Estimates are provided with the apparent order and theoretical order.}
\centering
\renewcommand{\arraystretch}{1.3}
\begin{tabular*}{\textwidth}{@{\extracolsep{\fill}}lccc} \hline  & $\langle n_{i,d} \rangle$ & $\langle T_{e,d} \rangle$ & $\langle T_{i,d} \rangle$ \\
\hline
Apparent $p_d$ & $0.07$ & $0.92$ & $0.78$ \\
$f_{0.5h}$ (Grid: $300 \times 800 \times 180$) & $66.5$ & $3.6$ & $4.8$ \\
$f_{h}$ (Grid: $150 \times 400 \times 180$) & $69.8$ & $6.8$ & $8.2$ \\
$f_{2h}$ (Grid: $75 \times 200 \times 180$) & $73.2$ & $12.9$ & $14.1$ \\[6pt]
\hline
Theoretical $p_d=2$ &  &  & \\
$f_{0.5h}$ (Grid: $300 \times 800 \times 180$) & $2.3$ & $1.0$ & $1.1$ \\
$f_{h}$ (Grid: $150 \times 400 \times 180$) & $9.2$ & $4.2$ & $4.4$ \\
$f_{2h}$ (Grid: $75 \times 200 \times 180$) & $15.3$ & $9.7$ & $9.6$ \\[6pt]
\hline
\end{tabular*}
\label{tab:functional_discretization_estimates}
\end{table}

\section{Conclusion}
\label{sec:conclusion}

In this work, we demonstrated the importance of accurately choosing numerical parameters in stellarator plasma edge simulations with EMC3-EIRENE. Characterization of numerical errors, while often overlooked, is needed to properly interpret predictive modeling and compare with experimental data. By applying methods developed for 2D plasma edge simulations, the statistical error, bias, and time integration error have been measured in 3D HSX simulations. The discretization error is only partially measured resulting from its complexity relative to the other error contributions. 

In particular, it was found that the asymptotic scaling laws for the statistical error, bias, and time integration error hold true in stellarator simulations. Simulations were shown to exhibit high iteration-to-iteration correlation for upstream plasma quantities compared to downstream plasma quantities. The statistical error was measured and compared with the commonly used volume averaged relative change metric, and it was found that the relative change metric systematically underestimates the statistical error. This highlights a critical flaw in EMC3-EIRENE analysis where simulations are often deemed converged, despite significant noise remaining in the solution. Post-process averaging was further emphasized as a computationally effective way of reducing this noise. The bias was found to be relatively small compared to the statistical error, although neutral quantities exhibit a larger bias than plasma quantities. Neutral quantities were further found to be approximately independent of EMC3's numerical time steps. Error resulting from the time steps was analyzed for plasma quantities and found that the dominant error differed in stellarator simulations compared to tokamak simulations, relating to the increased importance of cross-field transport in stellarators. The discretization error was partially analyzed for pressure-weighted average downstream quantities and found that just the radial and poloidal component of the error was already notable ($6\%-8\%$) for the main grid used in this study. The addition of the field-aligned component of the error, cell merging, and the effect of mapping loss is expected to substantially increase this estimation, and remains as future work. A more in depth overview of recommendations to limit the numerical error in future EMC3-EIRENE simulations is provided in Appendix~\ref{sec:appendix_B}.

These results emphasize the need for numerical error quantification in EMC3-EIRENE, as significant errors on relatively high resolution simulations were observed for the relatively small device HSX. Due to their increased size, large scale research and reactor-relevant devices will experience far greater numerical errors, and therefore must be quantified in predictive modeling. 


\section*{Acknowledgements}
\label{sec:acknowledgements}

This work was funded by the U.S. Department of Energy under
grant number DE-FG02-93ER54222. This work was also supported in part through the computational resources provided by the Center for High Throughput Computing at the University of Wisconsin--Madison (https://doi.org/10.21231/GNT1-HW21).

\appendix
\section{Error parameterization}
\label{sec:appendix_A}

Throughout this work, the numerical errors of EMC3-EIRENE have been analyzed and parameterized with respect to their theoretical model utilizing an ordinary least-squares method for the statistical error, Richardson-like extrapolation for the bias and time integration error, and Richardson extrapolation for the radial and poloidal component of the discretization error. The numerical errors on a simulation solution can be written as
\begin{equation}
\begin{split}
    \epsilon_{total\,num} = &\; \frac{A'_{s,E}}{\sqrt{P_{E}I}}R_{(r,\theta)}^a
    + \frac{A'_{s,S}}{\sqrt{P_{S}I}}R_{(r,\theta)}^a
    + \frac{A'_{s,N}}{\sqrt{P_{N}I}}R_{(r,\theta)}^a
    + \frac{A_{b,E}}{P_{E}}R_{(r,\theta)}^{2a}
    + \frac{A_{b,S}}{P_{S}}R_{(r,\theta)}^{2a} \\
    &+ \frac{A_{b,N}}{P_{N}}R_{(r,\theta)}^{2a}
    + A_{t,E\parallel}(\Delta t_{E\parallel})
    + A_{t,E\perp}(\Delta t_{E\perp})
    + A_{t,S}(\Delta t_{S})
    + \epsilon_{d}
    \label{eq:tot_error}
\end{split}
\end{equation}
where the first three terms on the right hand side represent the statistical error from each module, the next three terms represent the bias, the next three represent the time integration error, and the final term represents the discretization error. The constants $A$ characterize the magnitude of each error contribution and $a$ characterize the scaling of $\epsilon_s$ and $\epsilon_b$ across radial and poloidal grid refinements. Note that this scaling is expected to increase under additional field-aligned grid refinements. For the statistical error, we absorb the correlation time $\tau$ into the constant $A'_s = A_s \sqrt{\tau}$ in equation (\ref{eq:stat_error}).

For ease of interpretation and in accordance with \cite{ghoosNumericalAccuracyConvergence2020}, the numerical parameters are normalized with respect to a reference value to scale the constants. Dropping $\epsilon_d$ as it is not fully measured in this analysis, the total error equation can then be written as
\begin{equation}
\begin{split}
    \epsilon_{total\,num} = &\; \frac{A'^*_{s,E}\sqrt{P^*}}{\sqrt{P_{E}I}}R_{(r,\theta)}^a
    + \frac{A'^*_{s,S}\sqrt{P^*}}{\sqrt{P_{S}I}}R_{(r,\theta)}^a
    + \frac{A'^*_{s,N}\sqrt{P^*}}{\sqrt{P_{N}I}}R_{(r,\theta)}^a
    + \frac{A^*_{b,E}P^*}{P_{E}}R_{(r,\theta)}^{2a} \\
    &+ \frac{A^*_{b,S}P^*}{P_{S}}R_{(r,\theta)}^{2a}
    + \frac{A^*_{b,N}P^*}{P_{N}}R_{(r,\theta)}^{2a}
    + A^*_{t,E\parallel}\frac{\Delta t_{E\parallel}}{\Delta t^*}
    + A^*_{t,E\perp}\frac{\Delta t_{E\perp}}{\Delta t^*} \\
    &+ A^*_{t,S}\frac{\Delta t_{S}}{\Delta t^*}
    \label{eq:tot_error_normalized}
\end{split}
\end{equation}
with reference particles per iteration of $P^*=10^5$, and a reference time step of $\Delta t^*=5 \times 10^{-8}$ s. The normalized error constants are provided in table~\ref{tab:error_constants} for volumetric data along the field lines shown in figure~\ref{fig:field_lines} and for pressure-weighted average downstream plasma parameters. Only one digit is presented in this table as we only seek to represent the order of magnitude estimate of the errors. Note that these constants do not yet account for safety factors applied to error estimations. 

\begin{table}
\caption{Normalized constants $[-]$ for error contributions in (\ref{eq:tot_error_normalized}). '--' corresponds to estimates in which an error is negligible on a quantity. Constants are provided for field data and pressure-weighted average downstream plasma quantities.}
\centering
\renewcommand{\arraystretch}{1.3}
\begin{tabular*}{\textwidth}{@{\extracolsep{\fill}}lcccccccccc}
\hline
 & $n_i$ & $M$ & $T_e$ & $T_i$ & $n_H$ & $T_H$ & $p_{tot,n}$ & $\langle n_{i,d} \rangle$ & $\langle T_{e,d} \rangle$ & $\langle T_{i,d} \rangle$\\
\hline

$A'^*_{s,E}\times10^{3}$ & -- & -- & $100$ & $100$ & -- & -- & -- & $10$ & $7$ & $10$\\
$A'^*_{s,S}\times10^{3}$ & $40$ & $30$ & -- & -- & -- & -- & -- & $30$ & $5$ & $7$ \\
$A'^*_{s,N}\times10^{3}$ & -- & -- & -- & -- & $20$ & $30$ & $20$ & -- & -- & --\\
$A^*_{b,E}\times10^{3}$ & $4$ & $6$ & $8$ & $10$ & $6$ & $5$ & $5$ & $3$ & $70$ & $90$ \\
$A^*_{b,S}\times10^{3}$ & $0.9$ & $8$ & $0.6$ & $1$ & $5$ & $3$ & $5$ & $7$ & $1$ & $2$ \\
$A^*_{b,N}\times10^{3}$ & $2$ & $1$ & $0.7$ & $2$ & $90$ & $50$ & $70$ & $1$ & $0.9$ & $1$\\
$A^*_{t,E\parallel}\times10^{3}$ & $1$ & $3$ & $3$ & $2$ & -- & -- & -- & $4$ & $0.7$ & $0.5$ \\
$A^*_{t,E\perp}\times10^{3}$ & $1$ & $4$ & $3$ & $3$ & -- & -- & -- & $0.1$ & $3$ & $2$ \\
$A^*_{t,S}\times10^{3}$ & $0.1$ & $0.9$ & $0.1$ & $0.1$ & -- & -- & -- & $0.2$ & $0.009$ & $0.1$ \\
$a$ & $0.6$ & $0.4$ & $0.6$ & $0.6$ & $0.6$ & $0.4$ & $0.6$ & $0.4$ & $0.4$ & $0.4$ \\[6pt]
\hline
\end{tabular*}
\label{tab:error_constants}
\end{table}

\section{Recommendations to limit numerical errors in future EMC3-EIRENE studies}
\label{sec:appendix_B}

A significant challenge in EMC3-EIRENE simulations is finding a good balance between simulation accuracy and computational costs. In this section, some general statements to decrease the numerical errors are provided, alongside a discussion of the computational costs of simulations. These statements are based on findings presented in this study and those presented in \cite{ghoosNumericalAccuracyConvergence2020, ghoosAccuracyConvergenceIteratively2018}, but the goal is to make general statements on accuracy and best practices in EMC3-EIRENE.

It has been shown in this study that the volume averaged relative change metric (\ref{eq:relative_change}) can systematically underestimate the statistical error remaining on the simulation solution. Low relative change is a requirement of a steady solution but is not sufficient by itself in determining steadiness. It is critical for users to visibly monitor downstream plasma parameters and compare volumetric data fields between iterations to ensure a simulation is steady before taking a result. However, even if a simulation is determined to be steady by those metrics, significant statistical errors may remain in the solution. Post-process averaging is an effective tool to combat this remaining statistical noise and while it may appear to require greater computational effort, \cite{ghoosNumericalAccuracyConvergence2020} has shown that not to be the case. It is noted that the above recommendations still do not quantify deterministic errors remaining in the solution. 

It is the suggestion for all future simulations that users take $P_N$ at least a factor of two higher than $P_E$ and $P_S$. This gives users greater confidence that a steady solution is reached, despite EMC3-EIRENE providing no direct measure on the statistical error for neutral quantities. A measure of noise in neutral quantities should also be a future goal. This is especially relevant for high-recycling regimes and detached scenarios where plasma-neutral interactions play a more prominent role.

It further recommended that users take $\Delta t_{E,\parallel}=\Delta t_{E,\perp}=5\times10^{-8}$ s and $\Delta t_{S}=1\times10^{-7}$ s as default parameter choices. This ensures the error associated with $\Delta t_{E,\perp}$ doesn't significantly dominate the time integration error, which is the case in these HSX simulations if we take the default assumption of $\Delta t_{E,\perp} > \Delta t_{E,\parallel}$. For these simulations, the adjustment of the default value $\Delta t_{E,\perp}=2\times10^{-7}$ s to $\Delta t_{E,\perp}=5\times10^{-8}$ s was observed to increase the wall clock time of the ENERGY module by roughly $50\%$ ($100$ seconds). However, time is made back by increasing the default value of $\Delta t_{S}=5\times10^{-8}$ s to $\Delta t_{S}=1\times10^{-7}$ s, which reduced the wall clock time of the STREAMING module by roughly $30\%$ ($300$ seconds). In simulations with relatively smaller cross-field transport, one may take $\Delta t_{E,\perp} > \Delta t_{E,\parallel}$ without a loss to accuracy.

For the discretization error, although full quantification has not yet been performed, two points are clear. Firstly, the grid must be fine enough to resolve the gradients of the system. Secondly, the simulation solution must lie within the asymptotic regime in order to accurately measure the discretization error. However, the presence of cell merging has been shown to significantly complicate the confirmation of the asymptotic regime and the calculation of the discretization error. Therefore more skepticism must be placed on simulation results which do not quantify the discretization error. It is noted that agreement of the apparent order and the theoretical order is a strong indication that the asymptotic regime is reached, however disagreement does not necessarily indicate unsatisfactory results.

Users must also monitor the mapping loss associated with grid construction (subsection~\ref{subsec:discretization_error}) in order to ensure a significant amount of power is not lost. In practice, a power loss of $\leq 5 \%$ \cite{garciaExplorationResilientDivertors2025} is often deemed acceptable.

In general, users must themselves measure the numerical errors in their simulations. Although computationally inconvenient, users cannot strictly or confidently claim convergence (in the traditional CFD sense \cite{ProcedureEstimationReporting2008}) without the quantification of both the statistical and the deterministic numerical errors on their solution.

\bibliographystyle{iopart-num}
\bibliography{references}

\end{document}